\documentclass[12pt]{article}

\usepackage[margin=1in]{geometry}
\usepackage{natbib}
\usepackage{setspace}
\usepackage{hyperref}

\usepackage{amsmath,amssymb}
\newtheorem{theorem}{Theorem}

\newtheorem{corollary}{Corollary}
\newtheorem{lemma}{Lemma}

\usepackage{url}

\usepackage{graphicx} 

\usepackage{amsmath,bm}

\newcommand{\hide}[1]{}

\usepackage{float}
\usepackage{booktabs} 
\usepackage{multicol}
\usepackage{multirow}
\usepackage{comment}

\usepackage{thmtools}
\usepackage{thm-restate}
\usepackage{cleveref}

\def\X{\boldsymbol X}
\def\A{\boldsymbol A}
\def\by{\boldsymbol y}
\newcommand{\bbeta}{\boldsymbol{\beta}}
\def\bepsilon{\boldsymbol \epsilon}

\def\U{\boldsymbol E}
\def\bu{\boldsymbol e}

\def\tildeX{\boldsymbol Z}
\def\tildey{\tilde \by}

\def\widetildeG{\widetilde G}
\def\G{G}
\def\H{H}

\def\partialbeta{\tilde {\boldsymbol u}}
\def\partialbetaS{\tilde {\boldsymbol u}_S}
\def\partialbetaSC{\tilde {\boldsymbol u}_{S^c}}
\def\partialbetanobold{\tilde u}

\def\W{\boldsymbol W}
\def\h{\boldsymbol h}
\def\Y{\boldsymbol Y}

\def\gaussiantail{\delta_2}

\usepackage{algorithmic}
\usepackage[ruled, vlined]{algorithm2e}
\usepackage{soul}
\declaretheorem{assumption}

\begin{document}


\title{\bf Diffusion-Driven High-Dimensional Variable Selection}
\author{Minjie Wang\thanks{Department of Mathematics and Statistics, Binghamton University, State University of New York, Binghamton, NY},\hspace{.2cm}
Xiaotong Shen\thanks{School of Statistics, University of Minnesota, Minneapolis, MN},\hspace{.2cm}
and Wei Pan\thanks{Division of Biostatistics, University of Minnesota, Minneapolis, MN}}
\date{}
\maketitle

\begin{abstract}
Variable selection for high-dimensional, highly correlated data has long been a challenging problem, often yielding unstable and unreliable models. We propose a \emph{resample-aggregate} framework that exploits diffusion models' ability to generate high-fidelity synthetic data. Specifically, we draw multiple pseudo-data sets from a diffusion model fitted to the original data, apply any off-the-shelf selector (e.g., lasso or SCAD), and store the resulting inclusion indicators and coefficients. Aggregating across replicas produces a stable subset of predictors with calibrated stability scores for variable selection. Theoretically, we show that the proposed method is selection consistent under mild assumptions. Because the generative model imports knowledge from large pre-trained weights, the procedure naturally benefits from transfer learning, boosting power when the observed sample is small or noisy. We also extend the framework of aggregating synthetic data to other model selection problems, including graphical model selection, and statistical inference that supports valid confidence intervals and hypothesis tests. Extensive simulations show consistent gains over the lasso, stability selection, and knockoff baselines, especially when predictors are strongly correlated, achieving higher true-positive rates and lower false-discovery proportions. By coupling diffusion-based data augmentation with principled aggregation, our method advances variable selection methodology and broadens the toolkit for interpretable, statistically rigorous analysis in complex scientific applications. 
\end{abstract}

\noindent%

\newpage

\doublespacing

\section{Introduction}

Recent advances in generative models and generative artificial intelligence (AI) have prompted the advent of synthetic data generation, which challenges the traditional statistical practice of focusing solely on the raw data and shifts toward a more synthetic data-centric approach.
Synthetic data generation enjoys several advantages. First, it relieves data scarcity issues \citep{sufi2024addressing}. One of the major challenges for statistical analysis is insufficient data, which leads to difficulty in making accurate predictions or inferences. For example, insufficient samples may not provide enough statistical power to detect meaningful patterns in the data \citep{guo2013selecting}. Second, it addresses privacy concerns \citep{ghalebikesabi2023differentially} and is appealing in the scenarios where the real user data cannot be shared publicly while synthetic data are used as a substitute. For example, electronic health records data of patients, though useful resources for scientific discovery, are generally not publicly accessible due to privacy concerns.

Recently, denoising diffusion probabilistic models \citep{ho2020denoising,sohl2015deep} have become the key cornerstone of generative modeling and generative artificial intelligence.
Diffusion models aim to approximate the target distribution through forward and reverse Markov process where the forward process adds Gaussian noise to the initial sample from the data distribution while the reverse diffusion process gradually denoises a latent variable and allows generating new data samples.
It has gained increasing popularity due to its ability to generate high-resolution and high-fidelity images from the original ones.
Diffusion probabilistic models have been employed to generate synthetic data in many applications including image diffusion \citep{sohl2015deep}, text diffusion \citep{gong2022diffuseq}, time-series diffusion \citep{lin2024diffusion}, and tabular diffusion \citep{kotelnikov2023tabddpm}. In particular, \citet{kotelnikov2023tabddpm} developed a diffusion model (TabDDPM) that can be applied to the general tabular data case and handles features of any data type.

Further, transfer learning, which distills knowledge from large pre-trained models trained previously on large datasets from relevant studies, is known to elevate the accuracy of a deep learning model \citep{5288526}. Recent work has provided empirical evidence demonstrating the effectiveness of diffusion models augmented with transfer learning in generating high-quality image and tabular data \citep{shen2023boosting,tian2025conditional}.

On the other hand, variable selection is a central and long-standing problem in statistics, attracting substantial research interest. Among the most widely used approaches is the lasso, which introduces an $\ell_1$ penalty to encourage sparsity in high-dimensional linear regression models. Numerous extensions of the lasso have been developed to accommodate broader modeling contexts, including generalized linear models \citep{vandeGeer2008high}, fused lasso \citep{tibshirani2005sparsity}, and group lasso \citep{yuan2006model}. In graphical model selection, methods with similar $\ell_1$-type regularization have been proposed and studied extensively \citep{yuan2007model,banerjee2008model,friedman2008sparse}. To reduce the bias of the lasso, alternative nonconvex penalty functions have been proposed, such as the smoothly clipped absolute deviation (SCAD, \citeauthor{Fan01122001}, \citeyear{Fan01122001}) and minimax concave (MC+, \citeauthor{10.1214/09-AOS729}, \citeyear{10.1214/09-AOS729}) penalties.
In addition to penalty-based techniques, other classical methods for variable selection include univariate filtering, stepwise regression (both forward and backward), and best subset selection. However, despite their widespread adoption, these methods face critical limitations in modern high-dimensional settings. For example, when covariates are highly correlated, the lasso often violates the irrepresentable condition, resulting in unstable variable selection. Moreover, small sample sizes exacerbate these challenges, increasing the risk of false negatives. Crucially, traditional frequentist approaches often lack valid $p$-values for selected variables, precluding rigorous statistical inference or uncertainty quantification.

To address some of these shortcomings, stability selection \citep{meinshausen2010stability} has emerged as a robust alternative. By repeatedly applying a variable selection method to random subsamples of the data and selecting features with high selection frequency, stability selection offers improved robustness and interpretability. Many variants of this method have been studied, including an improved error control \citep{shah2013variable}. Parallel lines of work have focused on quantifying feature importance through model-based criteria. For example, random forests use the Gini index or permutation-based measures to evaluate a feature's contribution to predictive accuracy \citep{breiman2001random,altmann2010permutation}. Others have explored leave-one-variable-out strategies to assess predictive impact \citep{lei2018distribution,williamson2023general}.

In this paper, we propose a novel methodology for enhancing variable selection using synthetic data generated by diffusion models. These models faithfully replicate the distribution of the observed data and can be further refined via transfer learning from related pre-trained generative models when applicable. Our framework repeatedly applies a variable selection algorithm to multiple synthetic data sets to compute stability scores, as if we conducted repeated Monte Carlo simulations with a known data-generating distribution. Features are then selected based on their aggregated selection probabilities. On the other hand, we extend the framework of aggregating synthetic data to statistical inference to support valid hypothesis tests, analogous to the bootstrap, but driven entirely by synthetic data. Furthermore, our approach naturally accommodates transfer learning, allowing integration of information from large pre-trained diffusion models trained on related studies. Our core contribution lies in significantly improving the performance of classical Monte Carlo (MC) based techniques, such as bootstrapping and stability selection, through high-fidelity synthetic data generation. Unlike traditional approaches that rely on subsampling the observed data, we simulate and leverage multiple synthetic data sets from a learned generative model, capturing complex dependencies and enhancing generalization. In contrast to recent generative-model-based inference frameworks that use synthetic data solely to approximate null distributions of test statistics \citep{liu2024novel,shen2023boosting}, our method directly leverages MC simulations on synthetic data to compute feature stability scores for variable selection, closing the gap between generative modeling and stable variable selection.

\section{Methodology}

\subsection{Variable Selection}

Let $\by = (y_1,\cdots,y_n)^T$ be the vector of responses, $\X$ be the $n \times p$ design matrix, and $\bm \beta$ be a $p$-dimensional sparse vector.
Consider a regression model: $g(\mathbb E[\by| \X ]) = \X \bm\beta$ 
(e.g., linear model or generalized linear models) as an example and we demonstrate how to use synthetic data from diffusion models to perform variable selection. Our goal is to identify the set of informative features $\{j: \beta_j \neq 0\}$.

To perform variable selection, we compute the selection probability (or other feature importance metric) of each variable. A variable with a larger selection probability implies that it is more likely to be selected and hence significant in the model.

To compute the selection probability, we first train a diffusion model on the original data. Here, we utilize the diffusion model for tabular data (TabDDPM) proposed by \citet{kotelnikov2023tabddpm} for synthetic data generation.  Note that one can employ any other generative diffusion model for tabular data in the algorithm. 
We then use the trained diffusion model to generate $B$ synthetic data sets. The process can be viewed as if we conducted $B$ runs of Monte Carlo (MC) simulations with a known data-generating distribution.
In each simulation, we use one synthetic data set, apply a variable selection method (for example, the lasso) to it, and compute the indicator of whether a variable is selected by the variable selection method. We repeat this process until all $B$ synthetic data sets have been utilized.

\begin{algorithm}[ht!]
\SetKwInput{KwInput}{Input}
\SetKwInput{KwOutput}{Output}
\SetKwComment{Comment}{// }{}
\DontPrintSemicolon

\caption{Variable selection using synthetic data from diffusion models.}\label{model_selection}

\KwInput{Input data $\mathcal D = (\X_{\text{train}},\by_{\text{train}})$; A diffusion model generator $\mathcal G$; MC size $B$.}
\KwOutput{Estimated active set $\hat S$.}

Train a TABDDPM diffusion model $\mathcal G$ using $\mathcal D$.

\For{$b = 1,\cdots, B$}
{
Generate $\mathcal D^{(b)} = (\X_{\text{syn}}^{(b)},\by_{\text{syn}}^{(b)})$ using $\mathcal G$; 

Apply a variable selection method (for example, penalized $\ell_1$ regression with regularization parameter $\lambda^{(b)}$) to the synthetic data $\mathcal D^{(b)} = (\X_{\text{syn}}^{(b)},\by_{\text{syn}}^{(b)})$. 
Compute $I(\widehat \beta_j^{(b)} \neq 0)$, the indicator of whether the $j$th variable is selected from $\mathcal D^{(b)}$.

}

Compute selection probability: $\widehat \Pi_j = \frac{1}{B} \sum_{b=1}^B I(\widehat{\beta}_j^{(b)} \neq 0)$, $j = 1,\cdots,p$.

\Return Estimated active set $\hat S = \{j: \widehat \Pi_j \geq \pi_{\text{thres}} \}$.

\end{algorithm}

The selection probability for the $j$th variable is computed by averaging the selection indicators across all $B$ Monte Carlo simulations:
$$\widehat \Pi_j = \frac{1}{B} \sum_{b=1}^B I(\widehat{\beta}_j^{(b)} \neq 0),$$
which can be interpreted as the proportion of times variable $j$ is selected out of $B$ synthetic data sets (MC runs). For a true variable, $\widehat \Pi_j$ should be close to 1, while it should be much smaller and close to 0 for a noisy variable. Therefore, we declare the $j$th variable statistically significant if $\widehat \Pi_j \geq \pi_{\text{thres}}$ and obtain the set of selected stable features $\hat S = \{j: \widehat \Pi_j \geq \pi_{\text{thres}} \}$. In practice, the set of selected variables tends to be robust to a broad range of threshold choices $\pi_{\text{thres}}$.

To determine the optimal hyperparameters for the diffusion model, the regularization parameter $\lambda^{(b)}$ in the lasso, as well as the threshold level $\pi_{\text{thres}}$, we adopt the extended Bayesian Information Criterion (EBIC), which is particularly well-suited for high-dimensional settings where cross-validation is often unreliable and computationally expensive. Specifically, selecting the optimal $\lambda^{(b)}$ in the lasso on each synthetic data follows the standard EBIC approach; for a combination of diffusion hyperparameters and threshold level, we obtain the estimated active set $\hat S$, fit the ordinary least squares to the original data restricted on the set $\hat S$, and then compute the EBIC, following \citet{wang2024causal}. For low-dimensional scenarios where the irrepresentable condition typically holds, an alternative criterion based on the Fr\'echet Inception Distance (FID) between the original and synthetic data can also be used for hyperparameter tuning. We find that both selection strategies perform effectively in such settings and illustrate these results in Figure~\ref{fig:base_sim2}(B)  
and Figure~\ref{fig:sim_FID_F1score} in Appendix~\ref{sim_tune_FID}. However, in high-dimensional, highly correlated data regimes, the EBIC-based selection criterion consistently outperforms the FID approach in terms of variable selection accuracy.

The procedure is summarized in Algorithm~\ref{model_selection}. Importantly, Algorithm~\ref{model_selection} is broadly applicable to general model selection problems, including graphical model selection. Our framework can be interpreted as an ensemble method, analogous in spirit to the stability selection approach, in that it aggregates the results from multiple synthetic data sets to enhance the stability and reliability of variable selection. Further, instead of subsampling and using repeated samples, our approach leverages multiple, distinct, and high-quality synthetic data sets from a learned generative model to improve generalization and accuracy.

\subsection{Inference}

On the other hand, in the low-dimensional setting, our synthetic data generation framework can be applied to classical asymptotic statistical inference in terms of constructing standard errors, confidence intervals and statistical hypothesis testing (test statistics and $p$-values), similar to subsampling and bootstrapping. Our goal now is to test whether the regression coefficients are significant:
\begin{align*}
    H_0: \beta_j = 0, \quad H_a: \beta_j \neq 0.
\end{align*}
Recall that the test statistic for testing whether the $j$th regression coefficient is significant is: $T_j = {\widehat \beta_j} / {\text{SE}(\widehat \beta_j)}$, where $\text{SE}(\cdot)$ refers to the standard error of the estimator; the $p$-value of the test can be computed accordingly based on $T_j$.

Specifically, our approach can be applied to an arbitrary regression problem to facilitate hypothesis testing as follows. 
That is, we generate the synthetic data $B$ times and compute the $p$-value from each synthetic data set. The final estimated $p$-value is the average of $p$-values over $B$ synthetic data sets. We conclude that the $j$th variable is significant if its average $p$-value $\alpha_j$ is less than 0.05. We provide details in Algorithm~\ref{inference}.

\begin{algorithm}[ht!]
\SetKwInput{KwInput}{Input}
\SetKwInput{KwOutput}{Output}
\SetKwComment{Comment}{// }{}
\DontPrintSemicolon

\caption{Inference using synthetic data from diffusion models.}\label{inference}

\KwInput{Input data $\mathcal D = (\X_{\text{train}},\by_{\text{train}})$; A diffusion model generator $\mathcal G$; MC size $B$.}
\KwOutput{$p$-values indicating whether the regression coefficients are significant.}

Train a TABDDPM diffusion model $\mathcal G$ using $\mathcal D$.

\For{$b = 1,\cdots, B$}
{
Generate $\mathcal D^{(b)} = (\X_{\text{syn}}^{(b)},\by_{\text{syn}}^{(b)})$ using $\mathcal G$; 

Compute $T_j^{(b)} = {\widehat \beta_j^{(b)}} / {\text{SE}(\widehat \beta_j^{(b)})}$, $j = 1,\cdots,p$, where $\hat {\bm \beta}^{(b)}$ is the regression estimate from $\mathcal D^{(b)}$; compute $p$-value $\alpha_j^{(b)} = 2 \min(F_Z(T_j^{(b)}), 1 - F_Z(T_j^{(b)}))$, where $Z$ is the standard normal distribution.
}

\Return $p$-value $\alpha_j = \frac{1}{B} \sum_{b=1}^B \alpha_j^{(b)}$, $j = 1,\cdots,p$.

\end{algorithm}

\subsection{Knowledge Transfer}

Knowledge transfer has become increasingly important in generative modeling, as it allows models to leverage pre-trained knowledge from related domains to enhance generative accuracy and efficiency. This is particularly valuable when working with limited or noisy data. For example, a tabular diffusion model pre-trained on an Adult-Male dataset can be fine-tuned on an Adult-Female dataset with aligned features, enabling more effective learning through shared structures.

When relevant pre-trained models are available, we advocate for fine-tuning to facilitate knowledge transfer and incorporate inductive biases from prior related studies. This approach has been shown to improve statistical inference and variable selection, particularly in the context of synthetic data generation \cite{shen2023boosting}. By transferring knowledge from large, general-purpose models or domain-specific studies, one can achieve greater stability and accuracy in downstream tasks.

In cases where no suitable pre-trained model exists, training a generative model from scratch remains a viable and empirically effective strategy. Our numerical results confirm that diffusion-based models trained from scratch still yield strong performance, though transfer learning can further enhance outcomes when feasible.

\subsection{Theoretical Guarantees}

In this section, we establish the variable selection consistency of our proposed method in the context of linear models. Extensions to generalized linear models and graphical model selection follow similarly and are omitted for brevity.

We begin with the standard linear model:
\[
\by = \X \bm \beta^* + \bepsilon,
\]
where the observed synthetic data in the $b$th Monte Carlo (MC) sample is generated as
\[
\X_{\text{syn}}^{(b)} = \X + \U^{(b)}, \quad \by_{\text{syn}}^{(b)} = \by + \bu^{(b)}.
\]
Here, $\U^{(b)}$ and $\bu^{(b)}$ are the generalization errors introduced by the diffusion model. For notational simplicity, we denote $\X_{\text{syn}}^{(b)} = \tildeX^{(b)}$ and $\by_{\text{syn}}^{(b)} = \tildey^{(b)}$. The lasso is thus applied to $(\tildeX^{(b)}, \tildey^{(b)})$. We assume that the noise vector $\bepsilon$ is sub-Gaussian with variance proxy $\sigma^2$. Let $\bm \beta^*$ be a $p$-dimensional sparse vector with true support set $S := \mathrm{supp}(\bbeta^{*})= \{ j: \beta^*_j \neq 0 \}$ and $s:=|S|$.

We first impose two classical conditions required for the sign consistency of the lasso: the irrepresentable condition and the minimum eigenvalue condition on the population covariance matrix $\Sigma = n^{-1} \X^T \X$.

\begin{assumption}[Irrepresentable and Minimum Eigenvalue Conditions]
\label{irrepresentable_condition}
\[
\left \| {\Sigma}_{S^c, S} {\Sigma}_{S, S}^{-1} \right \|_{\infty} = 1 - \gamma < 1, \quad \text{and} \quad \Lambda_{\min}({\Sigma}_{S, S}) = C_{\min} > 0.
\]
\end{assumption}

Next, we assume that the synthetic data distribution closely approximates the distribution of the original data. Define:
\[
\widetilde{\Sigma}^{(b)} = \frac{1}{n} (\tildeX^{(b)})^T \tildeX^{(b)}, \quad \rho = \frac{1}{n} \X^T \by, \quad \tilde{\rho}^{(b)} = \frac{1}{n} (\tildeX^{(b)})^T \by.
\]

\begin{assumption}[Fidelity of Synthetic Data]
\label{closeness_condition}
There exists a parameter set $\theta$ characterizing the distribution of $\widetilde{\Sigma}^{(b)}$ and $\tilde{\rho}^{(b)}$. Then, for universal constants $C, c > 0$ and positive functions $\zeta$, $\varepsilon_0$ depending on $\theta$ and $\sigma^2$, the following tail bounds hold for all $\varepsilon \leq \varepsilon_0$:
\begin{align*}
\mathbb{P}\left(\left|\widetilde{\Sigma}^{(b)}_{ij} - \Sigma_{ij}\right| \geq \varepsilon \right) &\leq C \exp \left(-c n \varepsilon^{\nu} \zeta^{-1}\right), \quad \forall i,j = 1, \ldots, p, \\
\mathbb{P}\left(\left|\tilde{\rho}^{(b)}_j - \rho_j\right| \geq \varepsilon \right) &\leq C \exp \left(-c n s^{-2} \varepsilon^{\nu} \zeta^{-1}\right), \quad \forall j = 1, \ldots, p, \\
\mathbb{P}\left(\left| \frac{1}{n} \X_j^T (\tildey^{(b)} - \by) \right| \geq \varepsilon \right) &\leq C \exp \left(-c n s^{-2} \varepsilon^{\nu} \zeta^{-1} \right), \quad \forall j = 1, \ldots, p, \\
\mathbb{P}\left(\left| \frac{1}{n} (\tildeX_j^{(b)} - \X_j)^T (\tildey^{(b)} - \by) \right| \geq \varepsilon \right) &\leq C \exp \left(-c n s^{-2} \varepsilon^{\nu} \zeta^{-1} \right), \quad \forall j = 1, \ldots, p.
\end{align*}
\end{assumption}
Assumption~\ref{closeness_condition} essentially imposes tail bound conditions on the generalization errors of the diffusion model, which is substantiated by a growing body of recent work demonstrating the generalization capabilities of diffusion models \citep{li2023generalization,tian2024enhancing}.

We now establish the sign consistency of the lasso estimator $\widehat{\bbeta}^{(b)}$ fitted to synthetic data in a single MC iteration:
\[
\widehat{\bbeta}^{(b)} = \arg \min_{\bbeta} \left\{ \frac{1}{2n} \| \tildey^{(b)} - \tildeX^{(b)} \bbeta \|_2^2 + \lambda^{(b)} \| \bbeta \|_1 \right\}.
\]

\begin{theorem}[Uniform Selection Consistency for Synthetic Lasso]
\label{thm:consistency_each_syn}
Let $S=\operatorname{supp}(\bbeta^*)$ with $|S| = s$ and fix the regularization level $\lambda^{(b)}=\lambda$ for all synthetic replicates $b=1,\ldots,B$, where $\lambda \leq \min (\varepsilon_0, 4\varepsilon_0/\gamma)$; let $\varepsilon \leq \min(\varepsilon_1, \lambda / (\lambda \varepsilon_2 + \varepsilon_3))$, where the constants $\varepsilon_1, \varepsilon_2, \varepsilon_3 > 0$ depend on $\Sigma_{S,S}, \bbeta_S^*, \theta$, and $\sigma^2$. Suppose Assumptions~\ref{irrepresentable_condition} and~\ref{closeness_condition} hold.
Then, with probability at least
\[
1 - \delta_1, \quad \text{where } \delta_1 =  2p^2 C \exp \left(-c n s^{-2} \gamma^{\max(\nu,2)} \lambda^{\max(\nu,2)} \zeta^{-1} \right) + 2p^2 C \exp \left(-c n s^{-2} \varepsilon^{\max(\nu,2)} \zeta^{-1} \right),
\]
the following statements simultaneously hold for every Monte‑Carlo replicate $b=1,\ldots,B$:
\begin{enumerate}
\item[(a)] $\widehat{\bbeta}^{(b)}$ is unique and $\operatorname{supp}\bigl(\widehat{\bbeta}^{(b)}\bigr)\subseteq S$;
\item[(b)] $\bigl\|\widehat{\bbeta}^{(b)}_{S}-\bbeta^{*}_{S}\bigr\|_{\infty}\leq \kappa\lambda$, where $\kappa=6\|\Sigma_{S,S}^{-1}\|_{\infty}+C_{\min}^{-1/2}$;
\item[(c)] If $\min_{j\in S}|\bbeta^{*}_{j}|\geq \kappa\lambda$, then $\operatorname{sign}\bigl(\widehat{\bbeta}^{(b)}_{S}\bigr)=\operatorname{sign}\bigl(\bbeta^{*}_{S}\bigr)$.
\end{enumerate}
\end{theorem}
\begin{corollary}
    The estimate in a single MC iteration $\widehat{\bbeta}^{(b)}$ is sign-consistent if $\min_{j\in S}|\bbeta^{*}_{j}|\gg \kappa\lambda \gg \sqrt[{\max(\nu,2)}]{{(s^2 \log p)}/{n}}$ with probability tending to 1. When $\nu=2$, we obtain the standard rate of convergence of the lasso.
\end{corollary}

Theorem~\ref{thm:consistency_each_syn} ensures that variable selection using synthetic data is consistent in each MC simulation. As a result, the Monte Carlo estimate of the selection probability,
\[
\widehat \Pi_j = \frac{1}{B} \sum_{b=1}^B I(\widehat{\beta}_j^{(b)} \neq 0),
\]
tends to 1 for true variables and to 0 for null variables as $B$ and $n$ increase. Consequently, applying the union bound, with probability at least $1 - B \delta_1$, our selection algorithm correctly recovers the true support: $\widehat{S} = S$.

The proof of Theorem~\ref{thm:consistency_each_syn} is given in the Appendix. While our analysis builds upon the techniques of \citet{datta2017cocolasso}, we make two key theoretical contributions. First, unlike traditional measurement error model frameworks, we explicitly account for generalization errors introduced by diffusion-based synthetic data generation in both predictors and responses. Second, we relax the usual sub-Gaussian assumption to a broader sub-Weibull class by requiring only exponential tail bounds on the generalization error.
For simplicity, we present convergence rates under exponential tails, but our analysis shows that the consistency results extend to more general tail behaviors, such as polynomial bounds, under a suitable corresponding tail bound condition on the generalization error in Assumption~\ref{closeness_condition}.

\section{Simulation Studies}

This section investigates the performance of the proposed approach using synthetic data generation. All training and sampling activities utilize a single NVIDIA H100 GPU.

\subsection{Variable Selection and Graphical Model Selection}

We consider three statistical problems of model selection: sparse linear regression, sparse logistic regression, and graphical model selection. For sparse linear regression, we apply the lasso to the data; for sparse logistic regression, we use the $\ell_1$-regularized logistic regression; for graphical model selection, we use neighborhood selection.
For the sparse linear regression case, we consider three setups of the design matrix $\X$: i.i.d. design, auto-regressive design with $\Sigma_{ij} = \rho^{|i-j|}$ and $\rho=0.9$, and block-diagonal covariance structure design with blocks of size 5 and $\Sigma_{ij} = 0.9$, $\forall i \neq j$ within each block.
For the i.i.d. design, we set the number of non-zero coefficients of $\beta$ to be $s=5$ and the non-zero coefficients are randomly generated from \{-2,2\} plus small noise. For the auto-regressive design, we set $s=10$ and $\bbeta = (2,2,-2,-2,-2,2,2,-2,-2,-2,0,\cdots,0)$. For the block-diagonal design, we set $s=5$ and $\bbeta = (2,2,2,-2,-2,0,\cdots,0)$. For these two scenarios, we create a well-known challenging scenario where the data is highly correlated and the irrepresentable condition is violated so that the standard lasso method does not perform well.
For sparse logistic regression, we consider the same block-diagonal design $\X$ and simulate binary outcomes from Bernoulli distributions. 
For graphical model selection, we consider a small-world graph setup.
For all these base simulations, we fix the sample size to be $N=500$ and vary the number of features $p$.

We compare the variable selection accuracy obtained on the synthetic data against that obtained on the original raw data. 
For both approaches (applying the lasso to the raw or diffusion‑based synthetic data), tuning parameters are chosen via the extended BIC, ensuring a fair comparison. We set MC trials $B = 20$ and fix the synthetic sample size to equal the size of the training set; in Figure~\ref{fig:sim_tuneNsyn} of Appendix~\ref{sim_tune_syn_sample}, we treat the synthetic sample size as an additional hyperparameter and show that tuning the synthetic sample size yields performance that is close to simply setting it to the training-set size. Model selection performance is measured by the F1‑score, the harmonic mean of precision and recall, averaged over 10 trials.

\begin{figure}[ht]
    \centering
    \includegraphics[width = 1\textwidth]{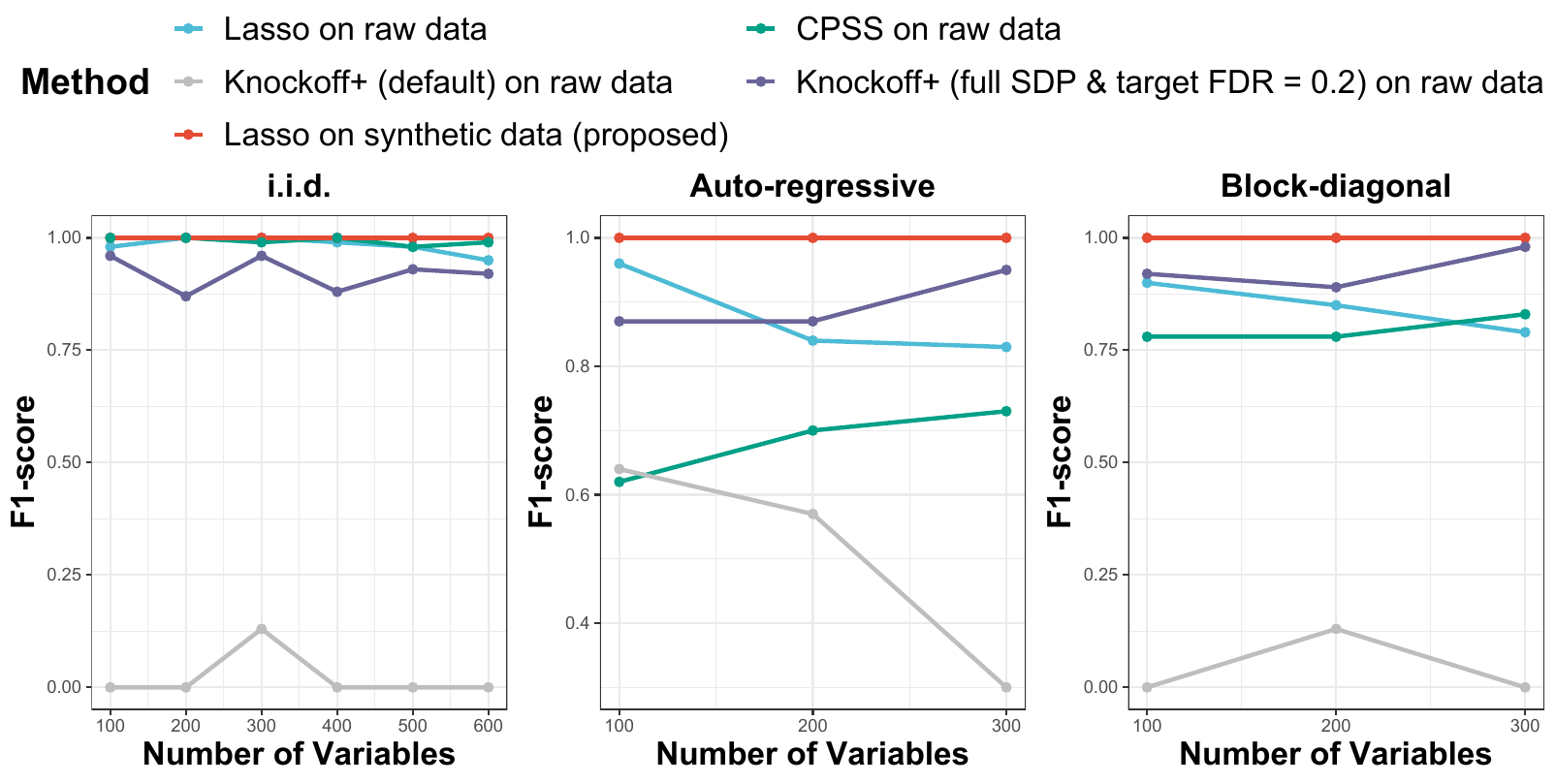}
    \caption{Variable selection accuracy (F1-score as the evaluation metric) for sparse linear regression. We compare the performance of applying lasso to the raw data (light blue) versus synthetic data (red), alongside complementary pairs stability selection (CPSS) and knockoff filters (default and full SDP with target FDR $= 0.2$). Three types of design matrices are considered: i.i.d., auto-regressive, and block-diagonal. The sample size is fixed at $N = 500$, while the number of features $p$ is varied across experiments.
}
    \label{fig:base_sim_lasso}
\end{figure}

Figure~\ref{fig:base_sim_lasso} shows that, under the i.i.d. design matrix, our approach applying the lasso to the synthetic data matches applying the lasso to the raw data when the number of variables $p$ is small; as $p$ grows, the accuracy of lasso on the raw data deteriorates, while our approach still performs well. For the challenging correlated designs, the advantage is even more pronounced: our method using synthetic data continues to perform well while running the lasso on raw data fails.

We further compare with two widely used competitors, stability selection and knockoffs \citep{barber2015controlling}.
For stability selection, we use its variant, complementary pairs stability selection (CPSS, \citeauthor{shah2013variable}, \citeyear{shah2013variable}) implemented in the \textsf{R} \texttt{stabs} package, which offers improved error control over the original procedure. 
For knockoffs, we employ the \textsf{R} \texttt{knockoff} package with (i) its default setting with approximate‑SDP construction (target $\operatorname{FDR}=0.1$) and (ii) the more powerful but computationally expensive full‑SDP construction with a relaxed target $\operatorname{FDR}=0.2$, to avoid zero‑selection runs in the i.i.d. and block-diagonal settings when using the default. Note that both competitors are sensitive to their hyperparameter choices, yet the number of true variables and the optimal FDR are unknown in practice.

Across all scenarios, our synthetic-data approach matches or exceeds the performance of CPSS and both knockoff configurations. Notably, CPSS captures more true positives as $p$ increases since its expected number of selected variables scales as $q = \sqrt{0.8 l p}$, while knockoffs yield fewer false positives for larger~$p$.

\begin{figure}
    \centering
    \includegraphics[width=0.9\linewidth]{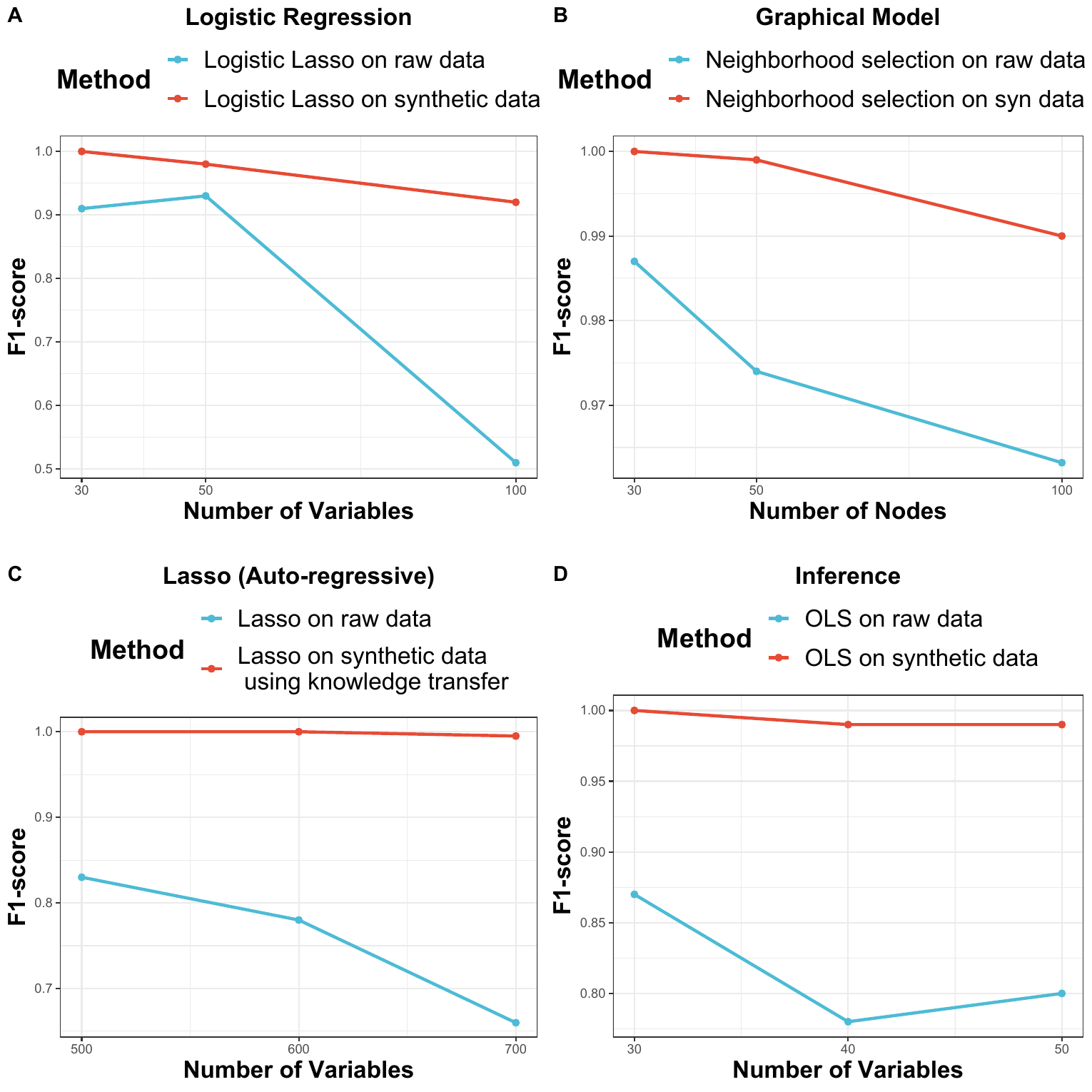}
    \caption{Simulation results comparing accuracy on raw data (blue) versus synthetic data (red) across four statistical tasks, using F1-score as the evaluation metric.
(A) \textbf{Sparse logistic regression}: $\ell_1$-penalized logistic regression with block-diagonal design matrix $\X$.
(B) \textbf{Graphical model selection}: Small-world network structure with $N = 500$ and varying $p$, using neighborhood selection.
(C) \textbf{Knowledge transfer in sparse linear regression}: Auto-regressive design with $N = 500$ and varying $p$; synthetic data generated via fine-tuned pre-trained model. Lasso is applied.
(D) \textbf{Statistical inference}: OLS with $N = 300$ and varying $p$; variables in raw data are deemed significant if $p$-value $< 0.05$.
}
    \label{fig:base_sim2}
\end{figure}

Moreover, Figure~\ref{fig:base_sim2} demonstrates that applying model selection methods to the synthetic data outperforms model selection on the raw data, for the task of sparse logistic regression (panel A) and graphical model selection (panel B). Even for a relatively large number of variables/nodes ($p = 100$), a challenging scenario where existing methods applied to raw data often fail, our method remains highly accurate.

In addition, to demonstrate that the synthetic data mimics the distribution of the raw data,
we show the pairwise correlation plots of the raw and synthetic data. We consider the graphical model selection setup and Figure~\ref{fig:sim_FID_correlation_plot} in Appendix~\ref{sim_tune_FID} indicates that the synthetic data retains the correlation structure of the raw data.

\subsection{Fine-tuning a Pre-trained Model}

In this section, we investigate the performance of variable selection on synthetic data generated from a model with knowledge transfer. Instead of training the diffusion model directly on the training data, we fine-tuned a pre-trained model, which leverages insights from previous similar studies with knowledge transfer. In particular, to achieve this, we generate a larger number of samples ($N = 2000$) from the same data-generating distribution and use them as pre-trained data. We first pre-trained the diffusion model on the pre-trained data and then further fine-tuned it on the training data. In addition, we also tune the sample size of the synthetic data here, allowing for synthetic volume expansion motivated by \citet{shen2023boosting}.

Here, we consider the variable selection task with auto-regressive design, as in the base simulation, but we create even more challenging scenarios by increasing the number of features $p$ to 500, 600, and 700. In these cases, the performance of training the diffusion model from scratch starts to decline as the generation error increases due to insufficient samples and large dimensions. Figure~\ref{fig:base_sim_lasso}(C) shows that applying variable selection to synthetic data generated from diffusion models with knowledge transfer still performs well.

\subsection{Inference}

In addition, we evaluate the performance of the proposed approach using synthetic data generation for the task of statistical inference. For the existing method, we apply ordinary least squares to the raw data and compute the test statistics from the regression model.
We determine that a variable is significant if its $p$-value is less than 0.05. For this simulation, we fix the sample size to be $N=300$ and vary the number of features $p$ from 30 to 50. 
Figure~\ref{fig:base_sim2}(D) shows the variable selection accuracy of inference on synthetic data compared with raw data.
Note that the traditional statistical inference method does not perform well as the dimension of the features increases. However, our proposed approach still performs well due to ensembles of multiple synthetic data sets.

\section{Case Study}

In this section, we apply the proposed synthetic data generation algorithm to the Alzheimer's Disease Neuroimaging Initiative (ADNI) dataset, publicly available at \url{https://adni.loni.usc.edu}. We focus on two statistical tasks: variable selection and causal discovery via DAG structure learning.

\subsection{Variable Selection}

First, our goal is to identify genes significantly associated with Alzheimer's disease (AD) status. Following the data preprocessing protocol of \citet{wang2024causal}, we first regress out potential confounding effects of additional covariates, including age, gender, education, handedness, and intracranial volume, from both the gene expressions and the binary AD status. The residuals from these regressions are used for further analysis. We perform marginal significance tests between each gene and the adjusted AD status, selecting those with $p$-values below $0.1$, resulting in $p = 30$ candidate genes.

We then apply the proposed algorithm using the lasso to identify relevant variables. The genes selected are summarized in Table~\ref{variable-selection-AD}. Several of these genes have been independently validated in the literature. For example, COX6B2 exhibits significantly increased mRNA expression in the hippocampal tissues of AD patients and mouse models \citep{bi2018genetic}. Similarly, \citet{zhao2016association} showed that COX8C is implicated in several neurodegenerative disorders, including Parkinson's, Huntington's, and Alzheimer's disease. CASP8 has also been linked to AD and is further identified in the causal analysis discussed below.

\begin{table}[ht]
	\vskip 0.05in
	\begin{center}
				\begin{tabular}{c}
					\toprule
					Selected Genes \\
					\midrule
					``CASP8",  ``COX6B2", ``COX8C",  ``ITPR3",  ``PPP3R2"  \\
					\bottomrule
				\end{tabular}
		\caption{Genes selected by the lasso using synthetic data generation.}
	\label{variable-selection-AD}	
	\end{center}
	\vskip -0.1in
\end{table}

\begin{figure}[ht!]
    \centering
    \includegraphics[width = 0.9\textwidth]{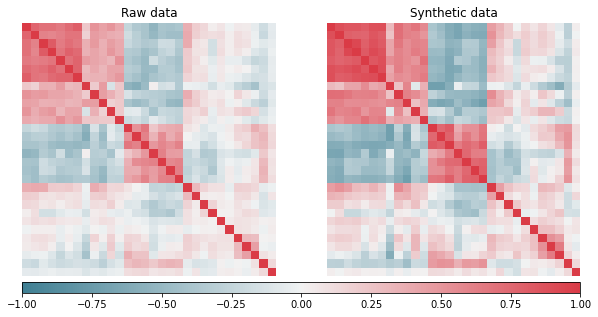}
    \caption{Pairwise correlation plots of the real and synthetic data for variable selection.}
    \label{fig:real_data_correlation_plot_regression}
\end{figure}

To evaluate fidelity, we visualize pairwise correlation plots of the real and synthetic data in Figure~\ref{fig:real_data_correlation_plot_regression}. The similarity between these plots confirms that the synthetic data preserves the correlation structure of the original dataset.

\subsection{DAG Structure Learning}

We next aim to estimate gene-to-gene and gene-to-AD regulatory relationships to identify causal effects among selected genes and disease status.

Following \citet{wang2024causal}, we first adjust gene expressions for confounders (age, gender, education, handedness, and intracranial volume) and retain the residuals. For each SNP corresponding to a gene, we conduct marginal significance tests against both gene expression and disease status. We select genes with at least one SNP whose $p$-value is less than $0.015$ with the gene and less than $0.02$ with disease status, yielding $p = 23$ primary variables. For each of these selected genes, we extract the SNP most strongly associated with AD as an instrumental variable, provided its $p$-value with the gene is below $0.015$, resulting in $q = 23$ instruments. We further include the disease status as an additional primary variable, along with its top-correlated SNP as an instrument, leading to $p = 24$ and $q = 24$ in total. Thus, we employ SNPs as instruments and gene expressions (along with disease status) as primary variables to infer the causal network.

To recover the DAG structure, we use \textsc{GAMPI}, a recently proposed method for causal discovery in high dimensions \citep{wang2024causal}. Section~\ref{sim_appendix_DAG} in the Appendix presents a simulation study verifying the effectiveness of synthetic data generation for DAG learning.

\begin{figure}[ht]
    \centering
    \includegraphics[width = 0.65\textwidth]{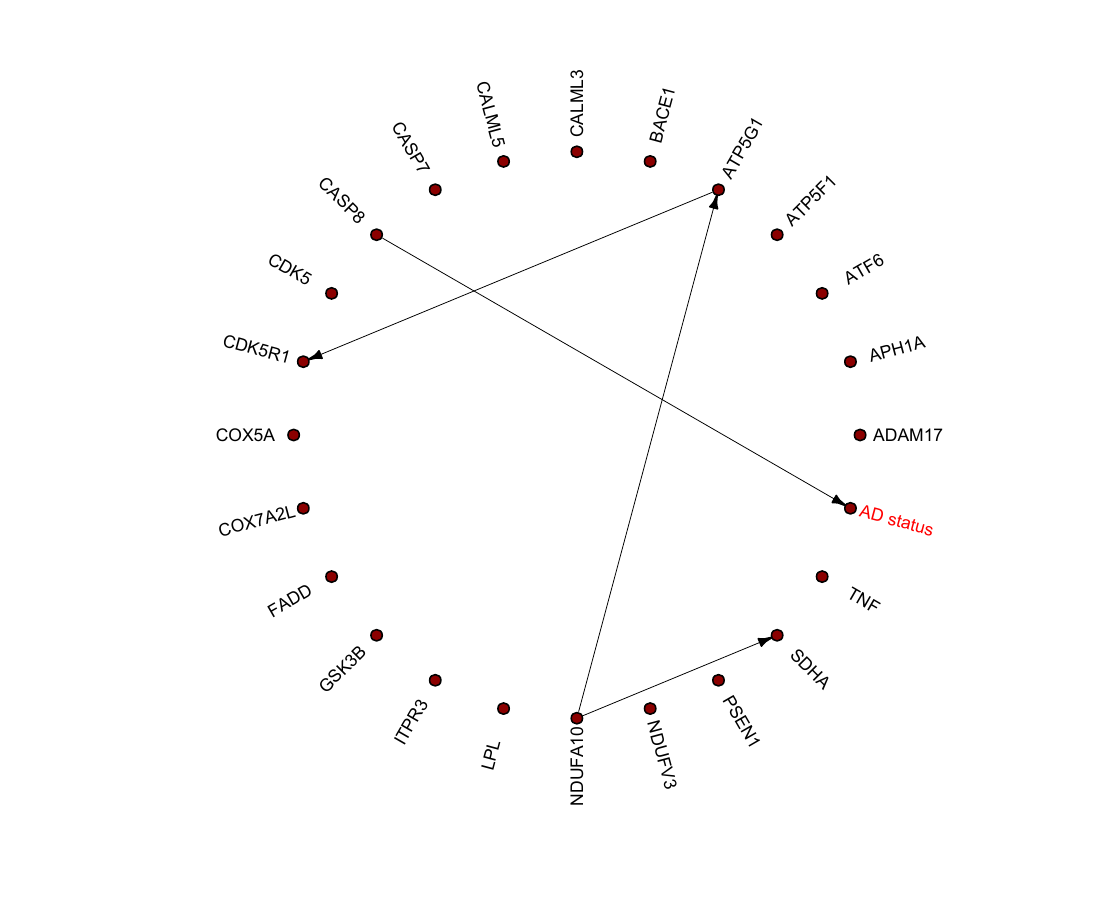}
    \caption{Reconstructed gene-to-gene and gene-to-AD regulatory network using synthetic data generation. ``AD status'' is a binary variable (0: normal, 1: AD). Arrows represent causal effects discovered by \textsc{GAMPI} applied to the synthetic data.}
    \label{fig:DAG_real_data}
\end{figure}

Figure~\ref{fig:DAG_real_data} shows the reconstructed network. Notably, CASP8 is identified as having a direct causal influence on AD status, consistent with known biological mechanisms. Caspases play a central role in neuronal apoptosis and neurodegeneration, with caspase-8 specifically linked to AD progression through cleavage of amyloid precursor proteins and increased amyloid-beta formation \citep{molecules25092071, CASP8}.

\begin{figure}[ht!]
    \centering
    \includegraphics[width = 0.9\textwidth]{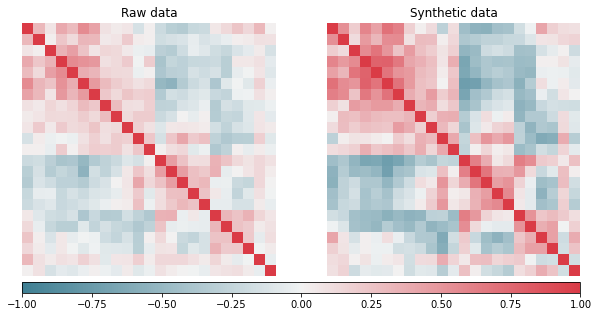}
    \caption{Pairwise correlation plots of real and synthetic data for DAG structure learning.}
    \label{fig:real_data_correlation_plot}
\end{figure}

To assess fidelity, we again compare pairwise correlation structures between real and synthetic data in Figure~\ref{fig:real_data_correlation_plot}. The strong similarity confirms that the diffusion model captures essential dependency structures.

\section{Discussion}

In this paper, we introduce a simple, general resample-aggregate framework that uses high-fidelity synthetic data generated by diffusion models to improve variable selection. In particular, we train a diffusion model on the original data, draw multiple synthetic data sets, apply a variable selection method to each, and compute feature scores by aggregating selection indicators across replicas. In this way, a stable subset of variables is selected. We also extend the framework of aggregating synthetic data to other model selection problems, including graphical model selection. We make theoretical contributions by demonstrating the selection consistency of the proposed approach under mild conditions on the generalization errors from the diffusion models. A series of empirical studies demonstrates that our approach outperforms existing variable selection methods for handling high-dimensional, highly correlated data.

In addition to the various statistical tasks exemplified in this paper, there are many possible directions for future research. For example, our framework can be further extended to other model selection problems, such as nonparametric regression and exponential family graphical model selection. The lasso is employed as the base selector in the algorithm, but other variable selection methods, such as SCAD and MC+, can be used.  While selection frequency is considered a measure, many other feature importance metrics, such as the Shapley value, can be adopted.

Overall, our work suggests that leveraging synthetic data achieves stable and accurate statistical performance for high-dimensional, highly correlated data and enhances generalization. We apply the proposed approach to Alzheimer's disease, yet it can be used to uncover underlying patterns in other complex problems with high-dimensional, highly correlated structures, such as neuroscience and public health. In conclusion, we develop a general, principled framework to address a challenging problem that yields strong empirical performance and opens many avenues for future research.

\section*{Acknowledgements}

This work was supported in part by NSF grant DMS-1952539, NIH grants R01GM113250, R01GM126002, R01AG065636, R01AG074858, R01AG069895, U01AG073079.

\newpage
\begin{appendix}

\begin{center}
{\bf \LARGE Diffusion-Driven High-Dimensional Variable Selection: Supplementary Materials} 
\bigskip

{\large Minjie Wang, Xiaotong Shen, and Wei Pan}
\end{center}

\section{Proof of Theorem~\ref{thm:consistency_each_syn}}

This section gives a complete proof of the consistency guarantees stated in Theorem~\ref{thm:consistency_each_syn}. Except for a few changes in notation (we remove the superscript
${}^{(b)}$ for ease of reading), the argument follows the classical \emph{primal-dual-witness} (PDW) construction of \citet{wainwright_2009_lasso} and its extension to measurement-error models by \citet{loh2012} and \citet{datta2017cocolasso}. Yet, unlike measurement error models, we explicitly account for generalization errors introduced by diffusion-based synthetic data generation in both predictors and responses. In addition, we relax the sub-Gaussian assumption to a broader sub-Weibull class by assuming exponential tail bounds for the generalization error. We nevertheless provide every step so that the paper is self‑contained.

\noindent\textbf{Remark.} Theorem~\ref{thm:consistency_each_syn} establishes the desired properties for a \emph{single} synthetic replicate $b$ and shows that the favorable event $\mathcal{E}^{(b)}$ occurs with probability at least $1-\delta_1$.
Because the $B$ replicates are generated independently, a union bound gives
\[
  \mathbb{P}\Bigl(\bigcap_{b=1}^{B}\mathcal{E}^{(b)}\Bigr)
  \;\geq\; 1-B\delta_1.
\]
Setting $\delta_B = B\delta_1$ proves the selection consistency of the proposed algorithm aggregating over $B$ replicates. The following proof of Theorem~\ref{thm:consistency_each_syn} proceeds exactly as in the single‑replicate case, and we retain the notation without the superscript $^{(b)}$ to avoid clutter.

\subsection*{Model and notation}
Let $(\tildeX,\tildey) = (\X_{\text{syn}}^{(b)},\by_{\text{syn}}^{(b)})$ be the synthetic data generated in the $b$‑th Monte Carlo replication and let
$ \by = \X\bbeta^{*}+\bepsilon $
with $\mathbb E[\bepsilon]=\mathbf 0$ and $\mathbb E[\bepsilon\bepsilon^{\!\top}]=\sigma^{2} \mathbf I_n$,
where $\mathbf I_n$ is the $n \times n$ identity matrix. Throughout the proof, we abbreviate the true support set $S := \mathrm{supp}(\bbeta^{*})= \{ j: \beta^*_j \neq 0 \}$ and $s:=|S|$.
For a matrix $\A$,  the sub‑matrix $\A_{S}$ corresponds to the subset of covariates that are in the support set $S$; $\A_{S,S}$ denotes the $s\times s$ sub‑matrix that only keeps the columns and rows indexed by~$S$. For a matrix $\A$, let $\| \A \|_{\infty} = \max_i \sum_j |a_{ij}|$ denote the matrix $\ell_{\infty}$ norm.

The synthetic design $\tildeX$ is an additive-noise version of the raw design $\X$,
$ \tildeX = \X + \U, $
where the rows of $\U$ are the generalization errors bounded by an exponential tail specified in the \emph{closeness condition} (Assumption~\ref{closeness_condition}). Similarly, the synthetic outcome $\tildey$ satisfies: $\tildey = \by + \bu$ where $\bu$ is the generalization error.

Moreover, throughout this section, following the proof of \citet{datta2017cocolasso}, $C$ and $c$ are denoted to be universal constants whose values
may vary across different expressions. We also introduce a few additional notations as follows.
\begin{align*}
& \Sigma = \frac{1}{n} \X^T \X, \quad \widetilde{\Sigma} = \frac{1}{n} \tildeX^T \tildeX, \\
& G = \Sigma_{S^c, S} \Sigma_{S, S}^{-1}, \quad \widetilde{G} = \widetilde{\Sigma}_{S^c, S} \widetilde{\Sigma}_{S, S}^{-1}, \quad H=\widetilde{G} - G, \quad F = \widetilde{\Sigma}_{S, S}^{-1} - \Sigma_{S, S}^{-1}, \\
& \phi = \left \|\Sigma_{S, S}^{-1} \right\|_{\infty}, \quad \psi= \left \|\Sigma_{S, S} \right\|_{\infty}, \quad B = \left \|\beta_S^* \right\|_{\infty} .
\end{align*}
Further, let $\delta_1(\varepsilon, \zeta) = p^2 C \exp \left(-c n s^{-2} \varepsilon^{\nu} \zeta^{-1}\right)$ and $\gaussiantail(\varepsilon, \zeta) = p^2 C \exp \left(-c n s^{-2} \varepsilon^2 \zeta^{-1}\right)$.

\subsection*{Step 1:  Primal-dual optimality conditions}

Let $\widehat{\bbeta}$ be the solution to the lasso problem:
\begin{align}
    \widehat{\bbeta} =  \arg \min_{\bbeta} \;\; \frac{1}{2n} \| \tildey - \tildeX \bbeta \|_2^2 + \lambda \| \bbeta \|_1 . \label{eq:lasso_problem}
\end{align}
The following lemma states the primal-dual optimality conditions.
\begin{lemma} \label{lemma:strict_dual_feasibility}
(a) A vector $\widehat \bbeta \in \mathbb{R}^p$ is the optimal solution to \eqref{eq:lasso_problem} if and only if there exists a subgradient vector $\partialbeta \in \partial |\widehat{\bbeta}|_1$ such that
\begin{align*}
    \frac{1}{n} \tildeX^T \tildeX \widehat{\bbeta} - \frac{1}{n} \tildeX^T \tildey  + \lambda \partialbeta & =0.
\end{align*}
(b) Suppose that the subgradient vector satisfies the strict dual feasibility condition $| \partialbetanobold_j | < 1$ for all $j \notin S(\widehat \bbeta)$. 
Then any other optimal solution $\widetilde{\bbeta}$ to the lasso satisfies $\widetilde \beta_j = 0$ for all $j \notin S(\widehat \bbeta)$.

\noindent (c) Under the conditions of part (b), if the $k \times k$ matrix $\tildeX_{S(\widehat{\bbeta})}^T \tildeX_{S(\widehat{\bbeta})}$ is invertible, then $\widehat{\bbeta}$ is the unique optimal solution of the lasso program.
\end{lemma}

\underline{Proof}: This lemma is a modified version Lemma 1 of \citet{wainwright_2009_lasso}. 
The proof is exactly analogous to that in the paper.

Following the PDW recipe, we construct a pair of candidate solution $(\widehat \bbeta^{\mathrm{PDW}},\partialbeta)$ that satisfies the KKT conditions with $\| \partialbetaSC \|_{\infty} < 1$; strict feasibility then implies uniqueness of the solution and $\mathrm{supp}
(\widehat\bbeta)\subseteq S$ by Lemma 2(a) of \citet{wainwright_2009_lasso}.

\subsection*{Step 2:  Primal-dual witness (PDW) construction}

Let $\widehat{\bbeta}_S$ be the solution to the restricted oracle lasso subproblem:
\begin{align*}
    \widehat{\bbeta}_S =  \arg \min_{\bbeta_S} \;\; \frac{1}{2n} \| \tildey - \tildeX_S \bbeta_S \|_2^2 + \lambda \| \bbeta_S \|_1 . 
\end{align*}

Let $\widehat{\bbeta} = (\widehat{\bbeta}_S^T, \mathbf 0_{(p-s) \times 1}^T)^T$ and $\partialbeta = (\partialbetaS^T,\partialbetaSC^T)^T$ where $\partialbetaS \in \partial (\| \widehat \bbeta_S \|_1)$ and $\partialbetaSC$ is some unspecified $(p-s)\times 1$ vector.
By part (a) of Lemma~\ref{lemma:strict_dual_feasibility}, we observe that $\widehat{\bbeta}$ is an optimal solution to \eqref{eq:lasso_problem} if and only if $(\widehat{\bbeta},\partialbeta)$ satisfies
\begin{align}
\frac{1}{n} \tildeX_S^T \tildeX_S \widehat{\bbeta}_S - \frac{1}{n} \tildeX_S^T \tildey  + \lambda \partialbetaS & =0, \label{eq:KKT_condition1}\\
\frac{1}{n} \tildeX_{S^c}^T \tildeX_S \widehat{\bbeta}_S - \frac{1}{n} \tildeX_{S^c}^T \tildey  + \lambda \partialbetaSC & =0 .  \label{eq:KKT_condition2}
\end{align}

By \eqref{eq:KKT_condition1}, 
\begin{align}
    \widehat{\bbeta}_S = (\frac{1}{n} \tildeX_S^T \tildeX_S)^{-1} ( \frac{1}{n} \tildeX_S^T \tildey  - \lambda \partialbetaS ) . \label{eq:betahat_expression}
\end{align}

\subsection*{Step 3:  Verifying strict dual feasibility (Theorem~\ref{thm:consistency_each_syn} Part (a))}

Plugging \eqref{eq:betahat_expression} into \eqref{eq:KKT_condition2} yields
\begin{align*}
    \partialbetaSC &= - \frac{1}{\lambda} \tildeX_{S^c}^T \tildeX_S (\tildeX_S^T \tildeX_S)^{-1} ( \frac{1}{n} \tildeX_S^T \tildey  - \lambda \partialbetaS ) + \frac{1}{\lambda} \cdot \frac{1}{n} \tildeX_{S^c}^T \tildey \\
    &= - \frac{1}{\lambda} \tildeX_{S^c}^T \tildeX_S (\tildeX_S^T \tildeX_S)^{-1} ( \frac{1}{n} \tildeX_S^T \by  - \lambda \partialbetaS ) + \frac{1}{\lambda} \cdot \frac{1}{n} \tildeX_{S^c}^T \by \; + \\
    & \quad\quad  \frac{1}{\lambda} \tildeX_{S^c}^T \tildeX_S (\tildeX_S^T \tildeX_S)^{-1} ( \frac{1}{n} \tildeX_S^T \by  - \frac{1}{n} \tildeX_S^T \tildey ) + \frac{1}{\lambda} \cdot \frac{1}{n} \tildeX_{S^c}^T (\tildey - \by) .
\end{align*}

By parts (b) and (c) of Lemma~\ref{lemma:strict_dual_feasibility}, we have that $\widehat{\bbeta}$ will be the unique solution to \eqref{eq:lasso_problem} if $\widetilde \Sigma_{S,S} = n^{-1} \tildeX_S^T \tildeX_S$ is nonsingular and $\|\partialbetaSC\|_{\infty} < 1$. 
By Lemma 6 of \citet{datta2017cocolasso}, $\mathbb P (\widetilde \Sigma_{S,S} \succ 0 ) \geq 1 - \delta_1(\varepsilon, \zeta)$ for all $\varepsilon \leq \min \left(\varepsilon_0, C_{\min} / 2\right)$.
We now establish the bounds for $\mathbb P \left( \|\partialbetaSC\|_{\infty} < 1 \right)$.
Recall we denote $\widetildeG = \tildeX_{S^c}^T \tildeX_S (\tildeX_S^T \tildeX_S)^{-1}$. Hence, we write $\partialbetaSC$ as
\begin{align*}
    \partialbetaSC &= \underbrace{\widetildeG \partialbetaS + \frac{1}{\lambda} ( \frac{1}{n} \tildeX_{S^c}^T \by - \widetildeG \frac{1}{n} \tildeX_S^T \by)}_{I_1} + \underbrace{\frac{1}{\lambda} \cdot \frac{1}{n} \tildeX_{S^c}^T (\tildey - \by)}_{I_2} + \underbrace{\frac{1}{\lambda} \cdot \frac{1}{n} \widetildeG \tildeX_S^T ( \by  - \tildey )}_{I_3} .
\end{align*}
We now bound $\|I_1\|_{\infty}$, $\|I_2\|_{\infty}$, and $\|I_3\|_{\infty}$ separately. The term $I_1$ has been analyzed by the proof of Theorem 2 Part (a) of \citet{datta2017cocolasso},
\begin{align*}
    I_1 = & \G \partialbetaS + \H \partialbetaS \\
    & + \frac{1}{\lambda}\left(\left(\frac{1}{n} \tildeX_{S^c}^T \by - \frac{1}{n} \X_{S^c}^T \by \right)+\left(\frac{1}{n} \X_{S^c}^T \by - \G \frac{1}{n} \X_{S}^T \by \right)+G\left( \frac{1}{n} \X_{S}^T \by - \frac{1}{n} \tildeX_S^T \by \right) - \H \frac{1}{n} \tildeX_S^T \by \right) \\
    = & \G \partialbetaS + \H \left( \partialbetaS + \frac{1}{\lambda}\left(\frac{1}{n} \X_{S}^T \by - \frac{1}{n} \tildeX_S^T \by \right)-\frac{1}{\lambda} \frac{1}{n} \X_{S}^T \by \right) \\
    & + \frac{1}{\lambda}\left(\left(\frac{1}{n} \tildeX_{S^c}^T \by - \frac{1}{n} \X_{S^c}^T \by \right) + \left(\frac{1}{n} \X_{S^c}^T \by - \G \frac{1}{n} \X_{S}^T \by \right) + G\left( \frac{1}{n} \X_{S}^T \by - \frac{1}{n} \tildeX_S^T \by \right)  \right) .
\end{align*}
Taking absolute values, we have
\begin{align*}
    \| I_1 \|_{\infty} &= \| \G \partialbetaS \|_{\infty} + \| \H \|_{\infty} \left( 1 + \frac{1}{\lambda} \| \frac{1}{n} \X_{S}^T \by - \frac{1}{n} \tildeX_S^T \by \|_{\infty} + \frac{1}{\lambda} \| \frac{1}{n} \X_{S}^T \by  \|_{\infty} \right) \\
    & + \frac{1}{\lambda} \| \frac{1}{n} \X_{S^c}^T \by - \G \frac{1}{n} \X_{S}^T \by \|_{\infty} + \left( \frac{1}{\lambda} \| \frac{1}{n} \tildeX_{S^c}^T \by - \frac{1}{n} \X_{S^c}^T \by  \|_{\infty} + \frac{1}{\lambda} \| G ( \frac{1}{n} \tildeX_S^T \by - \frac{1}{n} \X_{S}^T \by  ) \|_\infty \right).
\end{align*}
By the irrepresentable condition Assumption~\ref{irrepresentable_condition}, $\| \G \partialbetaS \|_{\infty} \leq 1 - \gamma$. In addition, for $\lambda \leq 4 \varepsilon_0 / \gamma$, we have
\begin{align*}
    & \mathbb P \left( \frac{1}{\lambda} \| \frac{1}{n} \tildeX_{S^c}^T \by - \frac{1}{n} \X_{S^c}^T \by  \|_{\infty} + \frac{1}{\lambda} \| G ( \frac{1}{n} \tildeX_S^T \by - \frac{1}{n} \X_{S}^T \by  ) \|_\infty < \gamma/2 \right) \\
    & \geq \mathbb P \left( \frac{1}{\lambda} \| \frac{1}{n} \tildeX^T \by - \frac{1}{n} \X^T \by  \|_{\infty} < \gamma/4 \right) \geq 1 - \delta_1(\lambda \gamma, \zeta).
\end{align*}
Moreover, by \cite{datta2017cocolasso}, $(\frac{1}{n} \X_{S^c}^T \by - \G \frac{1}{n} \X_{S}^T \by)$ is a linear combination of sub-Gaussian random variables, which yields
\begin{align*}
    \mathbb P \left( \frac{1}{\lambda} \| \frac{1}{n} \X_{S^c}^T \by - \G \frac{1}{n} \X_{S}^T \by \|_{\infty} \geq \gamma/4 \right) \leq \gaussiantail(\lambda \gamma, \zeta),
\end{align*}
where $\zeta$ is redefined as the maximum of the previous $\zeta$ and $\sigma^2$.
Finally, without loss of generality, we assume $\varepsilon_0 \leq 1$. By \cite{datta2017cocolasso}, with probability greater than $1-\delta_1(\varepsilon, \zeta) - \gaussiantail(\varepsilon, \zeta)$, $\| \frac{1}{n} \X_{S}^T \by - \frac{1}{n} \tildeX_S^T \by \|_{\infty} +  \| \frac{1}{n} \X_{S}^T \by  \|_{\infty} \leq \| \frac{1}{n} \X_{S}^T \by - \frac{1}{n} \tildeX_S^T \by \|_{\infty} +  \| \frac{1}{n} \X_{S}^T \bepsilon  \|_{\infty} + \| \frac{1}{n} \X_{S}^T \X_{S} {\bbeta}_S^* \|_{\infty} \leq 2 + B \psi$ for $\varepsilon \leq \min(1,\varepsilon_0)$. Then, with probability at least $1-\delta_1(\varepsilon, \zeta)- \gaussiantail(\varepsilon, \zeta)$,
\begin{align*}
    \| \H \|_{\infty} \left( 1 + \frac{1}{\lambda} \| \frac{1}{n} \X_{S}^T \by - \frac{1}{n} \tildeX_S^T \by \|_{\infty} + \frac{1}{\lambda} \| \frac{1}{n} \X_{S}^T \by  \|_{\infty} \right) \leq \left(1+\frac{1}{\lambda}(2+B \psi)\right) \frac{\varepsilon \phi(2-\gamma)}{(1-\phi \varepsilon)} \leq \frac{\gamma}{8},
\end{align*}
for $\varepsilon \leq \varepsilon_0^*$ where $\varepsilon_0^*=\min \left(\varepsilon_0, \gamma \lambda \phi^{-1}(8(2-\gamma)(\lambda+2+B \psi)+\gamma \lambda)^{-1}\right)$.
Combining all the probabilities, for $\lambda \leq 4 \varepsilon_0 / \gamma$ and $\varepsilon \leq \min(\varepsilon_0^*, C_{\min}/2)$, we have 
\begin{align*}
    \mathbb P( \| I_1 \|_{\infty} \geq 1 - \gamma / 8 ) \leq \delta_1(\lambda \gamma, \zeta) + \gaussiantail(\lambda \gamma, \zeta) + \delta_1(\varepsilon, \zeta) + \gaussiantail(\varepsilon, \zeta).
\end{align*}

For $I_2$, by Lemma~\ref{lemma:bound_product}, for $\lambda \leq 16 \varepsilon_0 / \gamma$,
\begin{align*}
    \mathbb P(\| I_2 \|_{\infty} \geq \frac{\gamma}{16} ) = \mathbb P ( \| \frac{1}{\lambda} \cdot \frac{1}{n} \tildeX_{S^c}^T (\tildey - \by) \|_{\infty} \geq \frac{\gamma}{16} ) \leq \delta_1(\lambda \gamma, \zeta) .
\end{align*}

For $I_3$, taking the absolute values, we have
\begin{align*}
    \| I_3 \|_{\infty} = \frac{1}{\lambda} \| \frac{1}{n} \widetildeG \tildeX_S^T (\tildey - \by) \|_{\infty} \leq \| \widetildeG \|_{\infty} \cdot \frac{1}{\lambda} \cdot \| \frac{1}{n}  \tildeX_S^T (\tildey - \by) \|_{\infty}.
\end{align*}

Note by the irrepresentable condition Assumption~\ref{irrepresentable_condition}, $\| G \|_{\infty} \leq 1$. Further, by Lemma 7 of \citet{datta2017cocolasso} and the triangular inequality,
\begin{align*}
    \| \widetildeG \|_{\infty} \leq \| G \|_{\infty} + \| \widetildeG - G \|_{\infty} =  \| G \|_{\infty} + \| H \|_{\infty} \leq 1 + \frac{\varepsilon \phi(2-\gamma)}{(1-\phi \varepsilon)}.
\end{align*}
Without loss of generality, we assume that $\varepsilon_0 \leq (2 \phi)^{-1}$.  Then with probability greater than $1-\delta_1(\varepsilon, \zeta)$, we have $\| \widetildeG \|_{\infty} \leq 1 + (2-\gamma)$ for $\varepsilon \leq \min (\varepsilon_0, (2 \phi)^{-1})$.

Meanwhile, similarly, by Lemma~\ref{lemma:bound_product},
\begin{align*}
    \| \frac{1}{n}  \tildeX_S^T (\tildey - \by) \|_{\infty} \leq \varepsilon,
\end{align*}
with probability greater than $1-\delta_1(\varepsilon, \zeta)$ for $\varepsilon \leq \varepsilon_0$. Therefore,
\begin{align*}
    \| I_3 \|_{\infty} = \frac{1}{\lambda} \| \frac{1}{n} \widetildeG \tildeX_S^T (\tildey - \by) \|_{\infty} \leq \left( 1 + (2-\gamma) \right) \cdot \frac{\varepsilon}{\lambda} \leq \frac{\gamma}{32},
\end{align*}
for $\varepsilon \leq \frac{\lambda \gamma}{32} \cdot ( 1 + (2-\gamma) )^{-1}$.

Combining the terms yields
\begin{align*}
    \| \partialbetaSC \|_{\infty} &\leq \|I_1\|_{\infty} + \|I_2\|_{\infty} + \|I_3\|_{\infty} \leq (1 - \gamma) + \left(\frac{\gamma}{2} + \frac{\gamma}{4} + \frac{\gamma}{8}\right) + \frac{\gamma}{16} + \frac{\gamma}{32} \\
    & = 1 - \frac{\gamma}{32},
\end{align*}
with probability at least $1 - \delta_1(\lambda \gamma, \zeta) - \gaussiantail(\lambda \gamma, \zeta) - \delta_1(\varepsilon, \zeta) - \gaussiantail(\varepsilon, \zeta)$ for $\lambda \leq 4 \varepsilon_0 / \gamma$ and $\varepsilon \leq \min ( \varepsilon_0^*, C_{\min} / 2, (2 \phi)^{-1} )$ where $\varepsilon_0^* = \min (\varepsilon_0, \gamma \lambda \phi^{-1}(8(2-\gamma)(\lambda + 2 + B \psi) + \gamma \lambda)^{-1},  \gamma \lambda ( 32( 1 + (2-\gamma) ) )^{-1} )$.
Therefore, $ \| \partialbetaSC \|_{\infty} < 1$, yielding strict dual feasibility.

\subsection*{Step 4: Establishing $\ell_{\infty}$ error bounds (Theorem~\ref{thm:consistency_each_syn} Part (b) and (c))}

By the expression of $\widehat{\bbeta}_S$ from \eqref{eq:betahat_expression}, we have
\begin{align*}
    \widehat{\bbeta}_S - {\bbeta}_S^* &= (\frac{1}{n} \tildeX_S^T \tildeX_S)^{-1} ( \frac{1}{n} \tildeX_S^T \tildey  - \lambda \partialbetaS ) - {\bbeta}_S^* \\
    &= (\frac{1}{n} \tildeX_S^T \tildeX_S)^{-1} ( \frac{1}{n} \tildeX_S^T \by  - \lambda \partialbetaS ) - {\bbeta}_S^*
    + (\frac{1}{n} \tildeX_S^T \tildeX_S)^{-1} ( \frac{1}{n} \tildeX_S^T (\tildey  - \by) ) \\
    &= \underbrace{(\frac{1}{n} \tildeX_S^T \tildeX_S)^{-1} ( \frac{1}{n} \tildeX_S^T \by -  \frac{1}{n} \X_S^T \by + \frac{1}{n} \X_S^T \X_S \bbeta_S^*+\frac{1}{n} \X_S^T \bepsilon - \lambda \partialbetaS ) - {\bbeta}_S^*}_{J_1} + \\
    & \quad \; \underbrace{(\frac{1}{n} \tildeX_S^T \tildeX_S)^{-1} ( \frac{1}{n} \tildeX_S^T (\tildey  - \by) )}_{J_2} .
\end{align*}
We now bound the $\ell_{\infty}$-norm of the above two terms separately. The term $J_1$ has been analyzed by the proof of Theorem 2 Part (b) and (c) of \citet{datta2017cocolasso},
\begin{align*}
    J_1 &= F_{S, S}\left( \frac{1}{n} \tildeX_S^T \by -  \frac{1}{n} \X_S^T \by + \frac{1}{n} \X_S^T \X_S \bbeta_S^*+\frac{1}{n} \X_S^T \bepsilon \right) \\
    & +\Sigma_{S, S}^{-1} \left( \frac{1}{n} \tildeX_S^T \by -  \frac{1}{n} \X_S^T \by \right)+\frac{1}{n} \Sigma_{S, S}^{-1} \X_S^T \bepsilon - \lambda \widetilde{\Sigma}_{S, S}^{-1} \partialbetaS ,
\end{align*}
where $F_{S, S} = \widetilde \Sigma_{S,S}^{-1} - \Sigma_{S,S}^{-1}$, $\Sigma_{S,S}^{-1} =  (\frac{1}{n} \X_S^T \X_S)^{-1}$ and $\widetilde \Sigma_{S,S}^{-1} = (\frac{1}{n} \tildeX_S^T \tildeX_S)^{-1}$.
By \citet{datta2017cocolasso}, following the proof of part (a), we conclude that for $\varepsilon \leq \varepsilon_0$, $\| \frac{1}{n} \tildeX_S^T \by -  \frac{1}{n} \X_S^T \by \|_{\infty} + \| \frac{1}{n} \X_S^T \X_S \bbeta_S^* \|_{\infty} +\| \frac{1}{n} \X_S^T \bepsilon \|_{\infty} \leq (2+B\psi)$ with probability at least $1 - \delta_1(\varepsilon, \zeta) - \gaussiantail(\varepsilon, \zeta)$. Therefore,
\begin{align*}
    \left \| F_{S, S}\left( \frac{1}{n} \tildeX_S^T \by -  \frac{1}{n} \X_S^T \by + \frac{1}{n} \X_S^T \X_S \bbeta_S^*+\frac{1}{n} \X_S^T \bepsilon \right) \right \|_{\infty} \leq \lambda \phi,
\end{align*}
with probability at least $1 - \delta_1(\varepsilon, \zeta) - \gaussiantail(\varepsilon, \zeta)$ for $\varepsilon \leq \lambda \phi^{-1}(\lambda+2+B \psi)^{-1}$; $\| \frac{1}{n} \tildeX_S^T \by -  \frac{1}{n} \X_S^T \by \|_{\infty} \leq \lambda$ with probability at least $1 - \delta_1(\lambda, \zeta)$ for $\lambda \leq \varepsilon_0$;
$\left\|\frac{1}{n} \Sigma_{S, S}^{-1} \X_S^T \bepsilon \right\|_{\infty} \leq \lambda / \sqrt{C_{\min }}$ with probability at least $1 - \gaussiantail(\lambda, \zeta)$; by Lemma 7 of \citet{datta2017cocolasso},
$\| \widetilde {\Sigma}_{S,S}^{-1} \|_{\infty} = \| (\frac{1}{n} \tildeX_S^T \tildeX_S)^{-1} \|_{\infty} \leq \phi + \| F \|_{\infty} \leq \phi+\phi^2 \varepsilon(1-\phi \varepsilon)^{-1} \leq 2 \phi$ with probability at least $1-\delta_1(\varepsilon, \zeta)$ for $\varepsilon \leq \min \left(\varepsilon_0,(2 \phi)^{-1}\right)$.
Therefore, combining these terms yields $\|J_1\|_{\infty} \leq \lambda (4 \phi + \frac{1}{\sqrt{C_{\min}}})$.

For the second term $J_2$, we have
\begin{align*}
    \| J_2 \|_{\infty} = \| (\frac{1}{n} \tildeX_S^T \tildeX_S)^{-1} ( \frac{1}{n} \tildeX_S^T (\tildey  - \by) ) \|_{\infty} \leq \| (\frac{1}{n} \tildeX_S^T \tildeX_S)^{-1} \|_{\infty} \cdot \| \frac{1}{n} \tildeX_S^T (\tildey  - \by) \|_{\infty} .
\end{align*}
Note $\widetilde{\Sigma}^{-1}_{S, S} =\Sigma^{-1}_{S, S} + F_{S, S}$ and $\phi = \| \Sigma_{S,S}^{-1} \|_{\infty}$. Again, by Lemma 7 of \citet{datta2017cocolasso}, we have with probability at least $1-\delta_1(\varepsilon, \zeta)$, for $\varepsilon \leq \min \left(\varepsilon_0,(2 \phi)^{-1}\right)$,
\begin{align*}
\| (\frac{1}{n} \tildeX_S^T \tildeX_S)^{-1} \|_{\infty} = \| \widetilde {\Sigma}_{S,S}^{-1} \|_{\infty} \leq \phi + \| F \|_{\infty} \leq \phi+\phi^2 \varepsilon(1-\phi \varepsilon)^{-1} \leq 2 \phi .
\end{align*}
Meanwhile, by Lemma~\ref{lemma:bound_product},
\begin{align*}
    \| \frac{1}{n} \tildeX_S^T (\tildey  - \by) \|_{\infty} \leq \lambda.
\end{align*}
with probability at least $1 - \delta_1(\lambda, \zeta)$ for $\lambda \leq \varepsilon_0$. Therefore, 
\begin{align*}
    \| J_2 \|_{\infty} = \| (\frac{1}{n} \tildeX_S^T \tildeX_S)^{-1} ( \frac{1}{n} \tildeX_S^T (\tildey  - \by) ) \|_{\infty} \leq 2 \lambda \phi .
\end{align*}

Combining all the probabilities, we have
\begin{align*}
    \| \widehat{\bbeta}_S - {\bbeta}_S^* \|_{\infty} & \leq \bigg( \left \| F_{S, S}\left( \frac{1}{n} \tildeX_S^T \by -  \frac{1}{n} \X_S^T \by + \frac{1}{n} \X_S^T \X_S \bbeta_S^*+\frac{1}{n} \X_S^T \bepsilon \right) \right \|_{\infty} \\
    & + \phi \| \frac{1}{n} \tildeX_S^T \by -  \frac{1}{n} \X_S^T \by \|_{\infty} + \| \frac{1}{n} \Sigma_{S, S}^{-1} \X_S^T \bepsilon\|_{\infty} + 2 \lambda \phi \bigg) + 2 \lambda \phi \\
    &\leq \lambda (4 \phi + \frac{1}{\sqrt{C_{\min}}} ) + 2 \lambda \phi \\
    & = \lambda (6 \phi + \frac{1}{\sqrt{C_{\min}}} ),
\end{align*}
with probability at least $1-\delta_1(\lambda, \zeta) -\gaussiantail(\lambda, \zeta) -\delta_1(\varepsilon, \zeta) -\gaussiantail(\varepsilon, \zeta)$ for $\varepsilon \leq\left(\varepsilon_0, C_{\min } / 2,(2 \phi)^{-1}, \lambda \phi^{-1}(\lambda+2+\right.$ $B \psi)^{-1}$ ) and $\lambda \leq \varepsilon_0$.

This proves part (b). If $| \bbeta_{\min}^*| > \lambda (6 \phi + \frac{1}{\sqrt{C_{\min}}} )$, then the lasso estimate is sign consistent.

\subsection*{Step 5:  Concluding the argument and probability of the high‑probability event}

In summary, by Step 3 and Lemma~\ref{lemma:strict_dual_feasibility}, the constructed
$(\widehat\bbeta,\partialbeta)$ is the unique optimal of the lasso program. This proves claims~(a) of the theorem. Step 4 proves part~(b) of the theorem. For part~(c),  if $| \bbeta_{\min}^*| = \min_{j\in S}|\bbeta^{*}_{j}| > \lambda (6 \phi + \frac{1}{\sqrt{C_{\min}}} )$, then the lasso estimate is sign consistent.

Combining the concentration bounds used in previous steps produces the event $\mathcal E$ on which every display above holds and
\[
    \mathbb P(\mathcal E^{c})
     \;\le\;
     2p^{2}C\exp\!\bigl(-c n s^{-2} \varepsilon^{\max(\nu,2)} \zeta^{-1}\bigr)
     \;+\;
     2p^{2}C\exp\!\bigl(-c n s^{-2} \lambda^{\max(\nu,2)} \gamma^{\max(\nu,2)} \zeta^{-1}\bigr)
     \;=\; O\!\bigl(p^{-1}\bigr),
\]
establishing the claimed exponential tail behavior.
$\blacksquare$

Last, we present a lemma required to prove Theorem~\ref{thm:consistency_each_syn}.
\begin{lemma}\label{lemma:bound_product}
Under the assumptions given in Assumption~\ref{closeness_condition}, 
the synthetic data $(\tildeX,\tildey)$ satisfy the following probability statement:
    \begin{align*}
        \mathbb{P}\left( \left| \frac{1}{n} \tildeX_{j}^T (\tildey - \by) \right| \geq \varepsilon\right) & \leq C \exp \left(-c n s^{-2} \varepsilon^2 \zeta^{-1}\right), \quad \forall j=1, \ldots, p  .
    \end{align*}
\end{lemma}

\underline{Proof}: By the triangular inequality,
\begin{align*}
    | \frac{1}{n} \tildeX_j^T (\tildey - \by) | \leq | \frac{1}{n} \X_j^T (\tildey - \by) | + | \frac{1}{n} (\tildeX_j - \X_j)^T (\tildey - \by) |.
\end{align*}
Therefore,
\begin{align*}
\mathbb{P} \left(\left | \frac{1}{n} \tildeX_j^T (\tildey - \by) \right| \geq \varepsilon\right) &\leq \mathbb{P} \left(\left | \frac{1}{n} \X_j^T (\tildey - \by) \right | + \left | \frac{1}{n} (\tildeX_j - \X_j)^T (\tildey - \by)  \right| \geq \varepsilon\right) \\
&\leq \mathbb{P} \left(\left | \frac{1}{n} \X_j^T (\tildey - \by) \right |  \geq \frac{\varepsilon}{2} \text{ or } \left | \frac{1}{n} (\tildeX_j - \X_j)^T (\tildey - \by)  \right| \geq \frac{\varepsilon}{2} \right) \\
&\leq \mathbb{P} \left(\left | \frac{1}{n} \X_j^T (\tildey - \by) \right |  \geq \frac{\varepsilon}{2} \right) 
+ \mathbb{P} \left(\left | \frac{1}{n} (\tildeX_j - \X_j)^T (\tildey - \by)  \right| \geq \frac{\varepsilon}{2} \right) \\
& \leq C \exp \left(-c n s^{-2} \varepsilon^2 \zeta^{-1}\right), \quad \forall j=1, \ldots, p .
\end{align*}

$\blacksquare$

\section{Additional Simulation}

\subsection{Tuning the Sample Size of Synthetic Data}\label{sim_tune_syn_sample}

\begin{figure}[ht!]
    \centering
    \includegraphics[width = 1\textwidth]{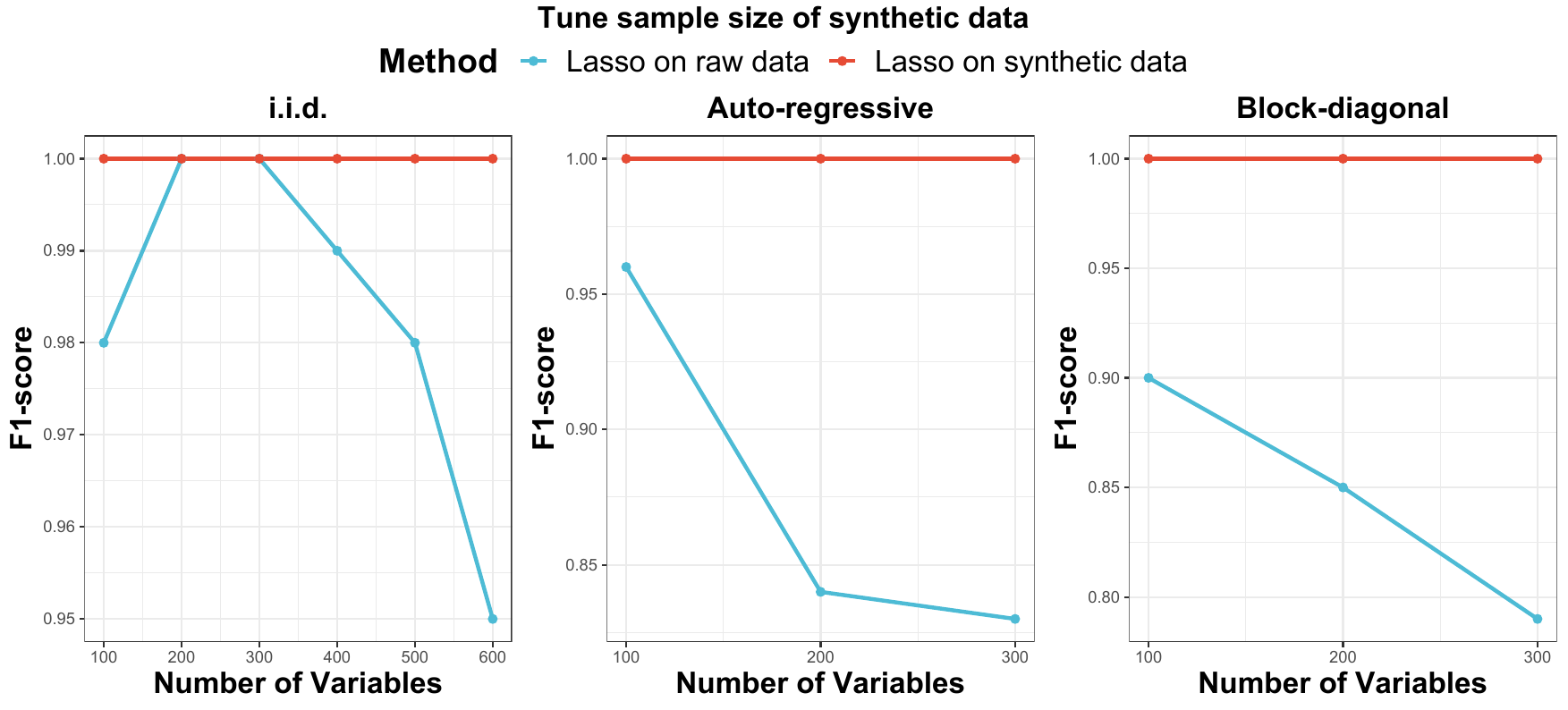}
    \caption{Simulation results for variable selection accuracy in sparse linear regression. We adopt the same simulation setup as in Figure~\ref{fig:base_sim_lasso} but tune the synthetic sample size instead of fixing it to be the number of training samples.}
    \label{fig:sim_tuneNsyn}
\end{figure}

In this experiment, we assess how the number of synthetic observations influences the model selection accuracy of our algorithm. In the base simulation, we fixed the synthetic sample size to equal the size of the training set. Here, we instead treat the synthetic sample size as an additional hyperparameter and, as in the main text, select its optimal value by minimizing the EBIC. We consider and revisit the sparse linear regression scenario used in Figure~\ref{fig:base_sim_lasso}. As reported in Figure~\ref{fig:sim_tuneNsyn}, choosing the synthetic sample size in this data-driven way yields performance that is nearly indistinguishable from simply setting it to the training-set size.
Consequently, in practice, we recommend setting the synthetic sample size equal to the number of training observations.

\subsection{Tuning Parameters with FID}\label{sim_tune_FID}

\begin{figure}[ht]
    \centering
    \includegraphics[width=0.5\linewidth]{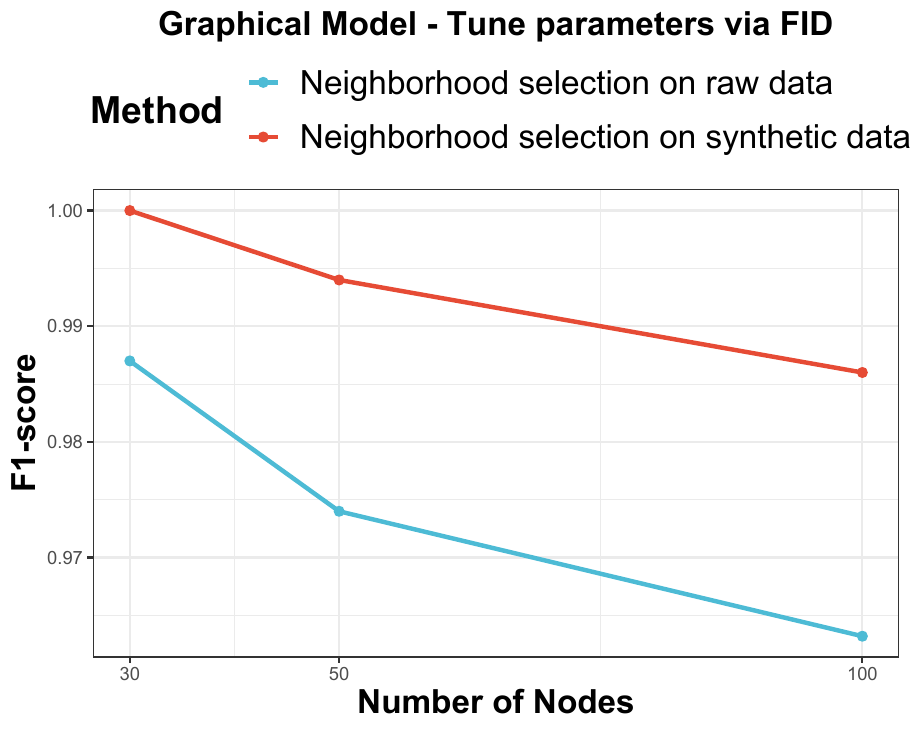}
    \caption{Edge selection accuracy for graphical models under the simulation setup of Figure~\ref{fig:base_sim2}(B). We compare the performance of applying neighborhood selection to the raw data (blue) with that to the synthetic data (red), where the diffusion-model hyperparameters are tuned by minimizing the Fr\'echet Inception Distance (FID).}
    \label{fig:sim_FID_F1score}
\end{figure}

In this experiment, we evaluate an alternative hyperparameter tuning strategy based on the Fr\'echet Inception Distance (FID). Recall that the base simulations choose hyperparameters by minimizing the EBIC; here, we instead minimize the FID between the observed data and the generated samples. Both criteria perform comparably on low-dimensional problems. Specifically, we consider and revisit the graphical model setting as in Figure~\ref{fig:base_sim2}(B). Figure~\ref{fig:sim_FID_F1score} demonstrates that FID-based hyperparameter tuning attains edge selection accuracy that is essentially indistinguishable from EBIC-based tuning.

\begin{figure}[ht]
    \centering
    \includegraphics[width=0.95\linewidth]{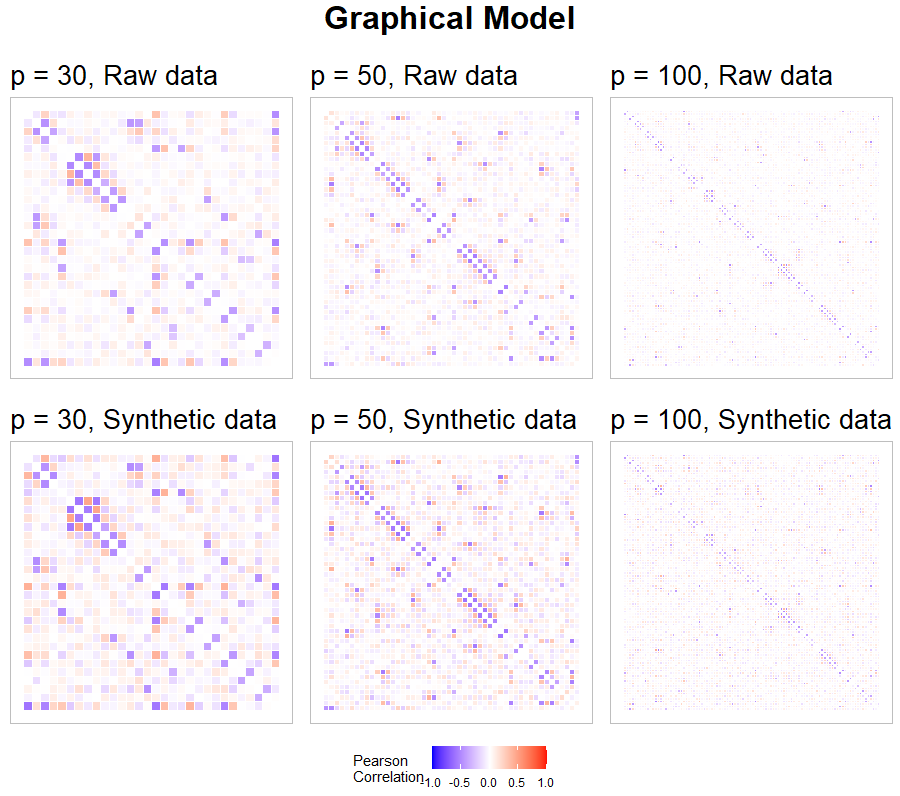}
    \caption{Pairwise correlation plots for the raw and synthetic data in the graphical model simulation with a small-world graph structure.}
    \label{fig:sim_FID_correlation_plot}
\end{figure}

We also examine how faithfully the synthetic data reproduces the joint distribution of the original variables. Figure~\ref{fig:sim_FID_correlation_plot} contrasts the pairwise correlation matrices of the observed and synthetic data, confirming that the diffusion model preserves the underlying correlation structure.

\subsection{DAG Structure Learning}\label{sim_appendix_DAG}

We next integrate our synthetic data generation scheme with a recently proposed DAG structure learning method \textsc{GAMPI} \citep{wang2024causal}.

\paragraph{Simulation design.}
We adopt the hub graph data-generating process of \citet{wang2024causal}.  
First, we create an adjacency matrix $\boldsymbol U$ with $U_{1j}=1$ for $j=2,\dots,p$ and zeros elsewhere, based on the hub graph structure, and construct an intervention matrix $\W$ whose diagonal entries satisfy $W_{jj}=1$ for $j=1,\dots,p$ while the off-diagonals satisfy $W_{lj}=0$ for $1\leq l\neq j\leq q$.

Secondly, for interventions, we generate binary instrumental variables $\X_{n \times q}= (\X_1,\ldots, \X_q)$ by first drawing an independent latent variable
$Z_{lj} \sim N(0, 1)$ and then setting \[
P(X_{lj}=1)=\frac{e^{Z_{lj}}}{1+e^{Z_{lj}}}, \quad  \text{and} \quad 
P(X_{lj}=0)=\frac{1}{1+e^{Z_{lj}}}.
\]
We further simulate confounders $\h \sim N(\mathbf{0},\boldsymbol{\Sigma})$ with $\rho_{ij}=0.1$.

Last, to mirror the setup of our real data analysis, we generate a graph of mixed data types, in which all the outcome variables $Y_j$ are Gaussian except the last, which is binary.  
For Gaussian outcomes, we set
\[
Y_j =
\begin{cases}
2\,\boldsymbol{w}_j^\top \X_{\mathrm{in}(j)} + h_j + \varepsilon_j, & \text{if }Y_j\text{ is a root variable},\\
0.5\,\boldsymbol{u}_j^\top \Y_{\mathrm{pa}(j)} + 3\,\boldsymbol{w}_j^\top \X_{\mathrm{in}(j)} + h_j + \varepsilon_j,& \text{otherwise.}
\end{cases}
\]
For the binary outcome, we draw $Y_j$ from a Bernoulli distribution with
\[
P(Y_j=1)=\frac{\exp\!\bigl(2\,\boldsymbol{u}_j^\top \Y_{\mathrm{pa}(j)} + 5\,\boldsymbol{w}_j^\top \X_{\mathrm{in}(j)} + h_j - 4.5\bigr)}
{1 + \exp\!\bigl(2\,\boldsymbol{u}_j^\top \Y_{\mathrm{pa}(j)} + 5\,\boldsymbol{w}_j^\top \X_{\mathrm{in}(j)} + h_j - 4.5\bigr)},
\]
where the intercept guarantees a balanced class distribution for the binary outcome.

\paragraph{Results.}
Table~\ref{sim_DAG} reports F1-scores for DAG reconstruction.  
Applying \textsc{GAMPI} to the synthetic data markedly outperforms applying it directly to the raw data. When the sample size is small, running \textsc{GAMPI} on the raw data tends to introduce spurious, small-magnitude entries in $\boldsymbol{V}$, which degrade DAG reconstruction. In contrast, our procedure generates multiple synthetic data sets, applies \textsc{GAMPI} to each, and aggregates the results through ensembling and thresholding, thereby eliminating these false positives.

\begin{table}[ht]
	\begin{center}
        \scalebox{1}{
			\begin{tabular}{lcc}
				\toprule
				\toprule
				\multirow{2}{*}{$(p,q,n)$} &
				\multicolumn{2}{c}{F1-score} \\ 
				    &   {Raw data} & {Synthetic data}  \\
				\midrule
				 (30,30,400) &   0.98 (5.9e-3) &   1.00 (0.00)  \\
				  (30,30,450) &   0.87 (9.7e-2) &   1.00 (0.00)  \\
				  (30,30,500) &    0.98 (7.7e-3) &    1.00 (0.00) \\
				\bottomrule
			\end{tabular}
   }
		\caption{DAG reconstruction accuracy of \textsc{GAMPI} under mixed outcomes. We compare its performance on the raw data with its performance on the synthetic data integrated with our algorithm.}
		\label{sim_DAG}
	\end{center}
\end{table}

\end{appendix}

\bibliographystyle{abbrvnat}
\bibliography{main.bib}

\begin{thebibliography}{37}
\providecommand{\natexlab}[1]{#1}
\providecommand{\url}[1]{\texttt{#1}}
\expandafter\ifx\csname urlstyle\endcsname\relax
  \providecommand{\doi}[1]{doi: #1}\else
  \providecommand{\doi}{doi: \begingroup \urlstyle{rm}\Url}\fi

\bibitem[Ahmad et~al.(2020)Ahmad, Sinha, Ahmad, Khalid, and Choi]{molecules25092071}
S.~S. Ahmad, M.~Sinha, K.~Ahmad, M.~Khalid, and I.~Choi.
\newblock Study of {C}aspase 8 inhibition for the management of {A}lzheimer's disease: A molecular docking and dynamics simulation.
\newblock \emph{Molecules}, 25\penalty0 (9), 2020.
\newblock ISSN 1420-3049.
\newblock \doi{10.3390/molecules25092071}.
\newblock URL \url{https://www.mdpi.com/1420-3049/25/9/2071}.

\bibitem[Altmann et~al.(2010)Altmann, Tolo{\c{s}}i, Sander, and Lengauer]{altmann2010permutation}
A.~Altmann, L.~Tolo{\c{s}}i, O.~Sander, and T.~Lengauer.
\newblock Permutation importance: a corrected feature importance measure.
\newblock \emph{Bioinformatics}, 26\penalty0 (10):\penalty0 1340--1347, 2010.

\bibitem[Banerjee et~al.(2008)Banerjee, Ghaoui, and d'Aspremont]{banerjee2008model}
O.~Banerjee, L.~E. Ghaoui, and A.~d'Aspremont.
\newblock Model selection through sparse maximum likelihood estimation for multivariate gaussian or binary data.
\newblock \emph{Journal of Machine Learning Research}, 9\penalty0 (15):\penalty0 485--516, 2008.

\bibitem[Barber and Cand{\`e}s(2015)]{barber2015controlling}
R.~F. Barber and E.~J. Cand{\`e}s.
\newblock {Controlling the false discovery rate via knockoffs}.
\newblock \emph{The Annals of Statistics}, 43\penalty0 (5):\penalty0 2055 -- 2085, 2015.
\newblock \doi{10.1214/15-AOS1337}.
\newblock URL \url{https://doi.org/10.1214/15-AOS1337}.

\bibitem[Bi et~al.(2018)Bi, Zhang, Zhang, Xu, Fan, Hu, Jiang, Tan, Li, Fang, et~al.]{bi2018genetic}
R.~Bi, W.~Zhang, D.-F. Zhang, M.~Xu, Y.~Fan, Q.-X. Hu, H.-Y. Jiang, L.~Tan, T.~Li, Y.~Fang, et~al.
\newblock Genetic association of the cytochrome c oxidase-related genes with {A}lzheimer's disease in {H}an {C}hinese.
\newblock \emph{Neuropsychopharmacology}, 43\penalty0 (11):\penalty0 2264--2276, 2018.

\bibitem[Breiman(2001)]{breiman2001random}
L.~Breiman.
\newblock Random forests.
\newblock \emph{Machine Learning}, 45:\penalty0 5--32, 2001.

\bibitem[Datta and Zou(2017)]{datta2017cocolasso}
A.~Datta and H.~Zou.
\newblock {CoCoLasso} for high-dimensional error-in-variables regression.
\newblock \emph{The Annals of Statistics}, 45\penalty0 (6):\penalty0 2400--2426, 2017.
\newblock \doi{10.1214/16-AOS1527}.

\bibitem[Fan and Li(2001)]{Fan01122001}
J.~Fan and R.~Li.
\newblock Variable selection via nonconcave penalized likelihood and its oracle properties.
\newblock \emph{Journal of the American Statistical Association}, 96\penalty0 (456):\penalty0 1348--1360, 2001.
\newblock \doi{10.1198/016214501753382273}.

\bibitem[Friedman et~al.(2008)Friedman, Hastie, and Tibshirani]{friedman2008sparse}
J.~Friedman, T.~Hastie, and R.~Tibshirani.
\newblock Sparse inverse covariance estimation with the graphical lasso.
\newblock \emph{Biostatistics}, 9\penalty0 (3):\penalty0 432--441, 2008.

\bibitem[Ghalebikesabi et~al.(2023)Ghalebikesabi, Berrada, Gowal, Ktena, Stanforth, Hayes, De, Smith, Wiles, and Balle]{ghalebikesabi2023differentially}
S.~Ghalebikesabi, L.~Berrada, S.~Gowal, I.~Ktena, R.~Stanforth, J.~Hayes, S.~De, S.~L. Smith, O.~Wiles, and B.~Balle.
\newblock Differentially private diffusion models generate useful synthetic images.
\newblock \emph{arXiv preprint arXiv:2302.13861}, 2023.

\bibitem[Gong et~al.(2022)Gong, Li, Feng, Wu, and Kong]{gong2022diffuseq}
S.~Gong, M.~Li, J.~Feng, Z.~Wu, and L.~Kong.
\newblock Diffuseq: Sequence to sequence text generation with diffusion models.
\newblock \emph{arXiv preprint arXiv:2210.08933}, 2022.

\bibitem[Guo et~al.(2013)Guo, Logan, Glueck, and Muller]{guo2013selecting}
Y.~Guo, H.~L. Logan, D.~H. Glueck, and K.~E. Muller.
\newblock Selecting a sample size for studies with repeated measures.
\newblock \emph{BMC Medical Research Methodology}, 13:\penalty0 1--8, 2013.

\bibitem[Ho et~al.(2020)Ho, Jain, and Abbeel]{ho2020denoising}
J.~Ho, A.~Jain, and P.~Abbeel.
\newblock Denoising diffusion probabilistic models.
\newblock \emph{Advances in Neural Information Processing Systems}, 33:\penalty0 6840--6851, 2020.

\bibitem[Jamal et~al.(2019)Jamal, Ali, Nagpal, Grover, and Grover]{CASP8}
S.~Jamal, W.~Ali, P.~Nagpal, A.~Grover, and S.~Grover.
\newblock Machine learning from molecular dynamics trajectories to predict {C}aspase-8 inhibitors against {A}lzheimer's disease.
\newblock \emph{Journal of Translational Medicine}, 10:\penalty0 780, 07 2019.
\newblock \doi{10.3389/fphar.2019.00780.}

\bibitem[Kotelnikov et~al.(2023)Kotelnikov, Baranchuk, Rubachev, and Babenko]{kotelnikov2023tabddpm}
A.~Kotelnikov, D.~Baranchuk, I.~Rubachev, and A.~Babenko.
\newblock {T}ab{DDPM}: Modelling tabular data with diffusion models.
\newblock In \emph{Proceedings of the 40th International Conference on Machine Learning}, volume 202, pages 17564--17579. PMLR, 2023.

\bibitem[Lei et~al.(2018)Lei, G'Sell, Rinaldo, Tibshirani, and Wasserman]{lei2018distribution}
J.~Lei, M.~G'Sell, A.~Rinaldo, R.~J. Tibshirani, and L.~Wasserman.
\newblock Distribution-free predictive inference for regression.
\newblock \emph{Journal of the American Statistical Association}, 113\penalty0 (523):\penalty0 1094--1111, 2018.

\bibitem[Li et~al.(2023)Li, Li, Zhang, and Bian]{li2023generalization}
P.~Li, Z.~Li, H.~Zhang, and J.~Bian.
\newblock On the generalization properties of diffusion models.
\newblock \emph{Advances in Neural Information Processing Systems}, 36:\penalty0 2097--2127, 2023.

\bibitem[Lin et~al.(2024)Lin, Li, Li, Li, and Gao]{lin2024diffusion}
L.~Lin, Z.~Li, R.~Li, X.~Li, and J.~Gao.
\newblock Diffusion models for time-series applications: a survey.
\newblock \emph{Frontiers of Information Technology \& Electronic Engineering}, 25\penalty0 (1):\penalty0 19--41, 2024.

\bibitem[Liu et~al.(2024)Liu, Shen, and Shen]{liu2024novel}
Y.~Liu, R.~Shen, and X.~Shen.
\newblock Novel uncertainty quantification through perturbation-assisted sample synthesis.
\newblock \emph{IEEE Transactions on Pattern Analysis and Machine Intelligence}, 2024.

\bibitem[Loh and Wainwright(2012)]{loh2012}
P.-L. Loh and M.~J. Wainwright.
\newblock High-dimensional regression with noisy and missing data: Provable guarantees with nonconvexity.
\newblock \emph{Annals of Statistics}, 40\penalty0 (3):\penalty0 1637--1664, September 2012.
\newblock \doi{10.1214/12-AOS1018}.
\newblock URL \url{https://doi.org/10.1214/12-AOS1018}.

\bibitem[Meinshausen and B{\"u}hlmann(2010)]{meinshausen2010stability}
N.~Meinshausen and P.~B{\"u}hlmann.
\newblock Stability selection.
\newblock \emph{Journal of the Royal Statistical Society Series B: Statistical Methodology}, 72\penalty0 (4):\penalty0 417--473, 2010.

\bibitem[Pan and Yang(2010)]{5288526}
S.~J. Pan and Q.~Yang.
\newblock A survey on transfer learning.
\newblock \emph{IEEE Transactions on Knowledge and Data Engineering}, 22\penalty0 (10):\penalty0 1345--1359, 2010.
\newblock \doi{10.1109/TKDE.2009.191}.

\bibitem[Shah and Samworth(2013)]{shah2013variable}
R.~D. Shah and R.~J. Samworth.
\newblock Variable selection with error control: another look at stability selection.
\newblock \emph{Journal of the Royal Statistical Society Series B: Statistical Methodology}, 75\penalty0 (1):\penalty0 55--80, 2013.

\bibitem[Shen et~al.(2023)Shen, Liu, and Shen]{shen2023boosting}
X.~Shen, Y.~Liu, and R.~Shen.
\newblock Boosting data analytics with synthetic volume expansion.
\newblock \emph{arXiv preprint arXiv:2310.17848}, 2023.

\bibitem[Sohl-Dickstein et~al.(2015)Sohl-Dickstein, Weiss, Maheswaranathan, and Ganguli]{sohl2015deep}
J.~Sohl-Dickstein, E.~Weiss, N.~Maheswaranathan, and S.~Ganguli.
\newblock Deep unsupervised learning using nonequilibrium thermodynamics.
\newblock In \emph{Proceedings of the 32nd International Conference on Machine Learning}, volume~37, pages 2256--2265. PMLR, 2015.

\bibitem[Sufi(2024)]{sufi2024addressing}
F.~Sufi.
\newblock Addressing data scarcity in the medical domain: A {GPT}-{B}ased approach for synthetic data generation and feature extraction.
\newblock \emph{Information}, 15\penalty0 (5):\penalty0 264, 2024.

\bibitem[Tian and Shen(2024)]{tian2024enhancing}
X.~Tian and X.~Shen.
\newblock Enhancing accuracy in generative models via knowledge transfer.
\newblock \emph{arXiv preprint arXiv:2405.16837}, 2024.

\bibitem[Tian and Shen(2025)]{tian2025conditional}
X.~Tian and X.~Shen.
\newblock Conditional data synthesis augmentation.
\newblock \emph{arXiv preprint arXiv:2504.07426}, 2025.

\bibitem[Tibshirani et~al.(2005)Tibshirani, Saunders, Rosset, Zhu, and Knight]{tibshirani2005sparsity}
R.~Tibshirani, M.~Saunders, S.~Rosset, J.~Zhu, and K.~Knight.
\newblock Sparsity and smoothness via the fused lasso.
\newblock \emph{Journal of the Royal Statistical Society Series B: Statistical Methodology}, 67\penalty0 (1):\penalty0 91--108, 2005.

\bibitem[van~de Geer(2008)]{vandeGeer2008high}
S.~A. van~de Geer.
\newblock High-dimensional generalized linear models and the lasso.
\newblock \emph{The Annals of Statistics}, 36\penalty0 (2):\penalty0 614--645, 2008.

\bibitem[Wainwright(2009)]{wainwright_2009_lasso}
M.~J. Wainwright.
\newblock Sharp thresholds for high-dimensional and noisy sparsity recovery using $\ell_{1}$-constrained quadratic programming (lasso).
\newblock \emph{IEEE Transactions on Information Theory}, 55\penalty0 (5):\penalty0 2183--2202, 2009.
\newblock \doi{10.1109/TIT.2009.2016018}.

\bibitem[Wang et~al.(2024)Wang, Shen, and Pan]{wang2024causal}
M.~Wang, X.~Shen, and W.~Pan.
\newblock Causal discovery with generalized linear models through peeling algorithms.
\newblock \emph{Journal of Machine Learning Research}, 25\penalty0 (310):\penalty0 1--49, 2024.

\bibitem[Williamson et~al.(2023)Williamson, Gilbert, Simon, and Carone]{williamson2023general}
B.~D. Williamson, P.~B. Gilbert, N.~R. Simon, and M.~Carone.
\newblock A general framework for inference on algorithm-agnostic variable importance.
\newblock \emph{Journal of the American Statistical Association}, 118\penalty0 (543):\penalty0 1645--1658, 2023.

\bibitem[Yuan and Lin(2006)]{yuan2006model}
M.~Yuan and Y.~Lin.
\newblock Model selection and estimation in regression with grouped variables.
\newblock \emph{Journal of the Royal Statistical Society Series B: Statistical Methodology}, 68\penalty0 (1):\penalty0 49--67, 2006.

\bibitem[Yuan and Lin(2007)]{yuan2007model}
M.~Yuan and Y.~Lin.
\newblock Model selection and estimation in the gaussian graphical model.
\newblock \emph{Biometrika}, 94\penalty0 (1):\penalty0 19--35, 2007.

\bibitem[Zhang(2010)]{10.1214/09-AOS729}
C.-H. Zhang.
\newblock {Nearly unbiased variable selection under minimax concave penalty}.
\newblock \emph{The Annals of Statistics}, 38\penalty0 (2):\penalty0 894 -- 942, 2010.
\newblock \doi{10.1214/09-AOS729}.

\bibitem[Zhao et~al.(2016)Zhao, Bai, Cheng, Tao, Wang, Liang, Yin, Hang, and Shang]{zhao2016association}
Q.-j. Zhao, S.-c. Bai, C.~Cheng, B.-z. Tao, L.-k. Wang, S.~Liang, L.~Yin, X.-y. Hang, and A.-j. Shang.
\newblock Association between chromosomal aberration of {COX8C} and tethered spinal cord syndrome: array-based comparative genomic hybridization analysis.
\newblock \emph{Neural Regeneration Research}, 11\penalty0 (8):\penalty0 1333--1338, 2016.

\end{thebibliography}

\end{document}